%% file: main.tex
\documentclass{SciPost}

\hypersetup{
    colorlinks,
    linkcolor={red!50!black},
    citecolor={blue!50!black},
    urlcolor={blue!80!black}
}

\usepackage{nicematrix}
\usepackage{amsmath}
\usepackage{siunitx}
\usepackage{array}
\usepackage{braket}
\usepackage{physics}
\usepackage{float}
\usepackage{graphicx}
\usepackage{fp}
\usepackage{calc}
\usepackage{listings}
\usepackage{minted}
\usepackage{layout}	
\usepackage{booktabs}
\usepackage{xspace}
\usepackage{xcolor}

\usepackage[]{overpic}
\usepackage[np, autolanguage]{numprint}
\usepackage{multicol}
\usepackage{comment}
\usepackage{subcaption}
\usepackage{hyphenat}

\usepackage{amssymb}
\usepackage{amsthm}
\usepackage{amsfonts}

\usepackage{pdfpages}
\usepackage[]{mdframed}

\theoremstyle{remark}

\definecolor{lstgray}{gray}{0.96}

\definecolor{code_bkg}{rgb}{0.96, 0.96, 0.96}
\setminted{
	linenos=true,
	autogobble,
    tabsize=4,
    obeytabs=true,
    frame=lines,
    framesep=2mm,
    fontsize=\footnotesize,
    breaklines=true,
    breakanywhere=true,
    bgcolor=code_bkg
}
\DeclareRobustCommand{\vec}[1]{ 				
	\ifthenelse{\equal{#1}{\omega} \OR \equal{#1}{\varphi} \OR \equal{#1}{\alpha} \OR \equal{#1}{\beta} \OR \equal{#1}{\chi} \OR \equal{#1}{\delta} \OR \equal{#1}{\varepsilon} \OR \equal{#1}{\phi} \OR \equal{#1}{\epsilon} \OR \equal{#1}{\gamma} \OR \equal{#1}{\eta} \OR \equal{#1}{\iota} \OR \equal{#1}{\kappa} \OR \equal{#1}{\lambda} \OR \equal{#1}{\mu} \OR \equal{#1}{\nu} \OR \equal{#1}{\pi} \OR \equal{#1}{\theta} \OR \equal{#1}{\vartheta} \OR \equal{#1}{\rho} \OR \equal{#1}{\sigma} \OR \equal{#1}{\varsigma} \OR \equal{#1}{\tau} \OR \equal{#1}{\upsilon} \OR \equal{#1}{\xi} \OR \equal{#1}{\psi} \OR \equal{#1}{\zeta}}{
		\boldsymbol{#1}
	}{
		\mathbf{#1}
	}
}

\usepackage[bitstream-charter]{mathdesign}
\DeclareSymbolFont{usualmathcal}{OMS}{cmsy}{m}{n}
\DeclareSymbolFontAlphabet{\mathcal}{usualmathcal}

\fancypagestyle{SPstyle}{
\fancyhf{}
\lhead{\colorbox{scipostblue}{\bf \color{white} ~SciPost Physics Codebases }}
\rhead{{\bf \color{scipostdeepblue} ~Submission }}

\fancyfoot[C]{\textbf{\thepage}}
}

\begin{document}

\pagestyle{SPstyle}

\begin{center}{\Large \textbf{\color{scipostdeepblue}{
FPGA-based LQG controller and hardware-in-the-loop simulator implementation for nanomechanical systems
}}}\end{center}

\begin{center}\textbf{
\renewcommand{\thefootnote}{\fnsymbol{footnote}}
Vojt\v{e}ch Mlyn\'{a}\v{r}\textsuperscript{1$\dagger$}\footnote[1]{These authors contributed equally to this work.},
Johannes Berndorfer\textsuperscript{1}\footnotemark[1],
Andreas Kugi\textsuperscript{1,2} and~Andreas~Deutschmann-Olek\textsuperscript{1$\ddagger$}
}\end{center}


\begin{center}
{\bf 1} Automation and Control Institute (ACIN), TU Wien, Vienna (1040), Austria
\\
{\bf 2} AIT Austrian Institute of Technology, Vienna (1210), Austria
\\[\baselineskip]
$\dagger$ mlynar@acin.tuwien.ac.at,\quad
$\ddagger$ deutschmann@acin.tuwien.ac.at
\end{center}

\section*{\color{scipostdeepblue}{Abstract}}
\textbf{\boldmath{%
We present an open-source framework for real-time Linear Quadratic Gaussian (LQG) control and hardware-in-the-loop (HIL) simulation on the affordable Red Pitaya STEMlab FPGA platform. The controller implements a discrete-time Kalman filter and Linear Quadratic Regulator (LQR) for systems with up to three coupled oscillatory degrees of freedom, targeting applications in levitated optomechanics, MEMS/NEMS, and related experimental platforms. Complementing the controller, the HIL simulator provides a~configurable second-order stochastic plant with nonlinear input and output mappings, enabling realistic closed-loop testing under real-time and fixed-point constraints. A MATLAB-based workflow automates model configuration, controller synthesis, numerical scaling, and FPGA deployment without requiring specialized hardware expertise. As an end-to-end demonstration, we present the stabilization of a levitated nanoparticle in a two-dimensional double-well potential, illustrating the complete workflow from model definition and simulation to real-time feedback control.
}}

\vspace{\baselineskip}

\noindent\textcolor{white!90!black}{%
\fbox{\parbox{0.975\linewidth}{%
\textcolor{white!40!black}{\begin{tabular}{lr}%
  \begin{minipage}{0.6\textwidth}%
    {\small Copyright attribution to authors. \newline
    This work is a submission to SciPost Physics Codebases. \newline
    License information to appear upon publication. \newline
    Publication information to appear upon publication.}
  \end{minipage} & \begin{minipage}{0.4\textwidth}
    {\small Received Date \newline Accepted Date \newline Published Date}%
  \end{minipage}
\end{tabular}}
}}
}


\vspace{10pt}
\noindent\rule{\textwidth}{1pt}
\tableofcontents
\noindent\rule{\textwidth}{1pt}
\vspace{10pt}


\section{Introduction}
\label{sec:intro}

Active feedback control and state estimation have become indispensable tools in levitated optomechanics \cite{millenOptomechanicsLevitatedParticles2020a,kremerAllelectricalCoolingOptically2024a,gieselerNonequilibriumSteadyState2015,liMillikelvinCoolingOptically2011,dagoStabilizingNanoparticlesIntensity2024,mlynarFeedbackStabilizationNanoparticle2026,magrini:2021}, nano- and micro-electromechanical systems (NEMS/MEMS) \cite{poggioFeedbackCoolingCantilevers2007, schmerlingOptimalSensingMomentum2025}, and other platforms that are strongly driven by external stochastic disturbances. In these settings, a state estimator paired with an optimal feedback law can suppress the resonant motion of a high-Q oscillator by several orders of magnitude or stabilize an otherwise unstable system. The optimal solution to this problem lies in the Linear Quadratic Gaussian (LQG) regulator: a Kalman filter (KF) for state estimation combined with a Linear Quadratic Regulator (LQR) for optimal control; it is mean-square-error optimal for the linear Gaussian plant model that describes most such oscillators in the regime of small displacements and extends directly to the quantum regime \cite{edwardsOptimalQuantumFiltering2005}. Another benefit is that its parameters follow from the physical noise covariances and cost weights without additional tuning parameters.

Despite the maturity of LQG theory, deploying it in a laboratory setting remains nontrivial. Natural frequencies in the tens to hundreds of kilohertz require sampling rates that exceed those of general-purpose computers and microcontrollers, necessitating the use of field-programmable gate arrays (FPGAs) or other specialized hardware. Additionally, these platforms introduce significantly greater complexity for the reliable implementation of such algorithms. Practical deployment, therefore, calls for an instrument that automates discretization, gain computation, fixed-point encoding, and model upload. While FPGA implementations of Kalman filters, LQR controllers, and feedback cooling systems have been reported in the literature, these solutions are typically tailored to a specific experiment, require substantial FPGA expertise, or are not available as openly accessible software packages. An integrated, openly available workflow that combines controller synthesis, hardware deployment, and real-time operation on low-cost hardware remains largely unavailable.

We present an integrated, open-source platform consisting of two complementary modules: an FPGA-based \textbf{LQG controller} and a \textbf{hardware-in-the-loop (HIL) simulator}, both designed for the Red Pitaya STEMlab 125-14 Z7020 and 125-14 Pro Z7020 Gen 2 boards, affordable, widely adopted instrument platforms built around a Xilinx Zynq Z7020 system-on-chip. A~MATLAB-based configuration workflow, backed by a lightweight onboard REST API server, lets the user input physical parameters, configure the controller, perform ADC and DAC calibration, and record internal states and signals at full sampling resolution through a graphical user interface, without requiring any embedded software or FPGA-development knowledge.

The LQG module supports systems with up to three coupled oscillatory degrees of freedom plus a drift term, covering the typical needs of optomechanical experiments. The instrument utilizes two 14-bit analog, DC-coupled voltage inputs and outputs, with a sampling rate of up to $\SI{125}{\mega\hertz}$. The FPGA implements the discrete-time Kalman filter and an LQR feedback law at a sampling period of $T_s = \SI{64}{\nano\second}$ using fixed-point arithmetic that balances precision with DSP resource consumption and throughput latency. The presented interface handles discretization, gain computation, and optional state-space normalizations that improve numerical conditioning on the fixed-point hardware. Two independently switchable parameter sets allow model switching during operation without halting the controller.

While the LQG implementation is, in principle, capable of representing any discretized well-scaled system with seven states, two inputs, and two outputs, the presented MATLAB workflow and GUI are designed for the specific model class of coupled harmonic oscillators with drift, described by the state-space equations in Section~\ref{sec:lqg}. This class covers a wide range
of experimental systems, but users requiring a different model structure would need to extend the configuration tooling.

The HIL simulator module implements a generalized plant model for stochastic, second-order systems with configurable separable nonlinearities in the damping, restoring force, and input/output mappings, driven by pseudorandom Gaussian white noise. It should be noted that the HIL simulator is not intended to serve as a precise twin of a real experiment, but rather as a tool that can emulate the behavior of such a system to a relatively high degree of fidelity. The primary aim is to provide a safe, low-consequence testing environment that can replace the actual plant in a feedback loop and expose the controller to a realistic signal environment. The simulator thus substantially reduces development risk by enabling controller validation before deployment on a physical experiment. A scenario-based MATLAB compilation step automatically determines the fixed-point scaling appropriate for the model and expected signal ranges.

As a concrete end-to-end demonstration, we configure and deploy both modules for a levitated nanoparticle exploring a two-dimensional double-well potential, reproducing the dynamics of our experimental work in~\cite{dagoStabilizingNanoparticlesIntensity2024,mlynarFeedbackStabilizationNanoparticle2026}. The example showcases both potential configurations, a fully confining quadratic potential and a saddle potential, and demonstrates live switching between them with the LQG controller active. The remainder of the article is organized as follows: Section~\ref{sec:lqg} covers the LQG controller's underlying mathematical model, cost function, fixed-point implementation, and configuration procedure; Section~\ref{sec:simulator} covers the HIL simulator architecture and model configuration workflow; and Section~\ref{sec:example} presents the end-to-end example with both boards in a closed feedback loop. This article primarily serves as a high-level overview of the implementation and a \textit{first-steps} guide; for in-depth technical references, we refer the reader to the documentation for both modules, available in their respective repositories.

\input{pages/lqg_sections.tex}

\input{pages/simulator_sections.tex}

\input{pages/example.tex}

\section{Conclusion}
In this work, we presented an open-source framework for real-time Linear Quadratic Gaussian (LQG) control and hardware-in-the-loop simulation on an accessible FPGA platform. The contribution combines two complementary modules: a real-time LQG controller and a hardware-in-the-loop simulator, both implemented for the Red Pitaya STEMlab 125-14 Z7020. Together, these tools address a practical gap between established control theory and routine laboratory deployment, where implementation complexity often prevents broader adoption.

We described the full controller workflow from model definition to hardware operation. Starting from a continuous-time state-space representation of coupled oscillatory dynamics, we detailed the discretization procedure, state estimation, and LQR gain synthesis used to design the controller. We then outlined implementation-specific elements required for reliable operation on FPGA hardware, including fixed-point arithmetic with per-element scaling and state-space transformations for consistent numerical behavior of the controller.

Beyond the controller itself, we introduced a simulator intended for realistic closed-loop testing under stochastic forcing and configurable nonlinear effects, while preserving the timing and numerical constraints of the target FPGA implementation. This module enables \linebreak scenario-based validation, parameter tuning, and safer experimentation before connecting to a physical plant, while still preserving relevant real-time and quantization constraints. The end-to-end example demonstrated how both modules can be configured, deployed, and switched between operating conditions, illustrating a practical pathway from model setup to live operation.

Our goal in developing and releasing this platform is to provide the community with a~transparent, extensible, and reproducible foundation for advanced feedback experiments. By lowering the technical barrier to high-speed state estimation and optimal control on low-cost hardware, we hope to support reproducible research and accelerate the adoption of advanced feedback techniques across levitated optomechanics, MEMS/NEMS, precision sensing, and related fields where real-time control is becoming increasingly important. Future work can build on this foundation by extending model classes, broadening hardware compatibility, and incorporating various additional features, such as a predictor for transport delay compensation or system identification procedures.

\section*{Acknowledgements}
This research was funded in part by the Austrian Science Fund (FWF) [10.55776/COE1, 10.55776/PAT9140723].

\bibliography{SciPost_Example_BiBTeX_File.bib}

\end{document}

%% file: pages/lqg_sections.tex
\section{LQG controller for nanomechanical systems}\label{sec:lqg}
This section presents the design and implementation of a Linear Quadratic Gaussian (LQG) controller tailored for nanomechanical systems. The LQG controller is a stochastically optimal approach that combines optimal state estimation via a Kalman filter with optimal feedback control via a Linear Quadratic Regulator. In the following sections, we detail the continuous-time mathematical formulation, a discretization scheme adapted to the FPGA sampling rate, and the fixed-point arithmetic strategies employed to maintain numerical stability and minimize resource consumption on the target hardware.

\subsection{Mathematical description}
To design the LQG controller, we first establish a mathematical model of the plant dynamics and measurement process and define a cost function that captures the control objectives. We model the plant as a linear continuous-time system with additive process and measurement noise, which is a common approximation for many nanomechanical systems, such as levitated nanoparticles and NEMS, operating near their equilibrium points. Specifically, we consider a system of three uncoupled harmonic oscillators with an additional drift term, and with optional cross-coupling in the input and output functions. This model structure was chosen because it captures the dominant dynamics of many levitated optomechanical and nanomechanical systems, including the three translational degrees of freedom typically encountered in such experiments, while maintaining a compact state-space representation that can be efficiently implemented on resource-constrained FPGA hardware. The additional drift state accounts for slow parameter variations or colored noise, and couplings between modes can still be represented through the input and measurement matrices.

Since the controller is to be deployed on an FPGA with a fixed sampling rate, we discretize the continuous-time model using standard techniques. The resulting discrete-time model serves as the basis for deriving the Kalman filter for state estimation and the LQR for optimal control presented in Section~\ref{subsec:lqg_implementation}.

We emphasize that the front-end interface is designed for the most common streamlined model structure, but the underlying implementation can support arbitrary linear state-space models with 7 states, allowing users to implement different systems by modifying only the MATLAB interface.

\subsubsection{System model}
The controller is designed to support three uncoupled oscillators (dubbed $x$, $y$, and $z$) and an additional drift term $\phi$ assembled in the state vector $\boldsymbol{\xi}$ as
\begin{equation}\label{eq:plant_ss_vec}
  \boldsymbol{\xi} = 
  \begin{bmatrix}
  x & \dot{x} & y & \dot{y} & z & \dot{z} & \phi \\
  \end{bmatrix}^\mathrm{T},
\end{equation}

The additional drift state $\phi$ is intended to capture slow baseline variations, low-frequency disturbances, or slowly varying offsets that cannot be represented by the oscillator dynamics themselves. In many practical experiments, this state improves estimator robustness in the presence of long-term drifts. The state-space model of a linear continuous-time system is described by
\begin{equation}\label{eq:plant_ss}
    \dot{\boldsymbol{\xi}}(t) = \mathbf{A}\boldsymbol{\xi}(t) + \mathbf{B}\mathbf{u}(t) + \mathbf{G}\mathbf{w}(t),
\end{equation}
where $\mathbf{u}\in\mathbb{R}^2$ collects the two input voltages, The use of two inputs is dictated by the two analog output channels available on the Red Pitaya STEMlab platform targeted by the present implementation. Moreover, $\mathbf{A}$ is the dynamic matrix 
\begin{equation}
  \mathbf{A} = \begin{bmatrix}
        0 & 1 &  &  &  &  &  \\
        -\mathrm{sgn}\left({\Omega_x}\right)\left|\Omega_x\right|^2 & -\Gamma_x &  &  &  &  &  \\
         &  & 0 & 1 &  &  &  \\
         &  & -\mathrm{sgn}\left({\Omega_y}\right)\left|\Omega_y\right|^2 & -\Gamma_y &  &  &  \\
         &  &  &  & 0 & 1 &  \\
         &  &  &  & -\mathrm{sgn}\left({\Omega_z}\right)\left|\Omega_z\right|^2 & -\Gamma_z &  \\
         &  &  &  &  &  & -1/\tau_\phi
    \end{bmatrix}
\end{equation}
with natural frequency $\Omega_i$ (potentially negative to model an unstable mode), damping rate $\Gamma_i$ with $i \in \{x, y, z\}$, and drift time constant $\tau_\phi$.
The matrices $\mathbf{G}$ and $\mathbf{B}$ feed the process noise $\mathbf{w}(t)=\left[w_x,w_y,w_z,w_{\phi}\right]^\mathrm{T}$, and the voltages $\mathbf{u}$, respectively, into the system through
\begin{equation}
  \mathbf{G} = 
  \begin{bmatrix}
    0 & 0 & 0 & 0 \\
    \frac{1}{m} & 0 & 0 & 0 \\
    0 & 0 & 0 & 0 \\
    0 & \frac{1}{m} & 0 & 0 \\
    0 & 0 & 0 & 0 \\
    0 & 0 & \frac{1}{m} & 0 \\
    0 & 0 & 0 & 1\\
  \end{bmatrix}, \quad 
  \mathbf{B} = \frac{1}{m}\begin{bmatrix}
      c_{f1,x} & c_{f2,x} \\
      c_{f1,dx} & c_{f2,dx} \\
      c_{f1,y} & c_{f2,y} \\
      c_{f1,dy} & c_{f2,dy} \\
      c_{f1,z} & c_{f2,z} \\
      c_{f1,dz} & c_{f2,dz} \\
      c_{f1,\phi} & c_{f2,\phi}
    \end{bmatrix},
\end{equation}
with the mass of the resonator $m$ and actuator coupling coefficients $c_f$. It should be noted that the mass-dependent scaling is optional, and the user can choose to set $m=1$ and absorb the mass into the other terms if desired.
Each process noise source is modeled as zero-mean, mutually independent white noise with covariance
\begin{equation}
    \mathbb{E}[\mathbf{w}(t)\mathbf{w}^\mathrm{T}(s)] = \mathbf{Q}\,\delta(t-s),\ \ \mathrm{where}\ \ \mathbf{Q}=\mathrm{diag}(\sigma_{w_x}^2,\sigma_{w_y}^2,\sigma_{w_z}^2,\sigma_{w_{\phi}}^2).
\end{equation}
The individual noise intensities allow the model to represent different disturbance intensities acting on each degree of freedom and on the drift state.

Measurements performed on the system yield two voltage outputs collected in
\begin{equation}\label{eq:obs_eq}
  \boldsymbol{\chi}(t) = \mathbf{C}\,\boldsymbol{\xi}(t) + \mathbf{v}(t),
\end{equation}
where the measurement matrix $\mathbf{C}$ reads as
\begin{equation}
  \mathbf{C} = \begin{bmatrix} c_{e1,x} & c_{e1,dx} & c_{e1,y} & c_{e1,dy} & c_{e1,z} & c_{e1,dz} & c_{e1,\phi} \\
  c_{e2,x} & c_{e2,dx} & c_{e2,y} & c_{e2,dy} & c_{e2,z} & c_{e2,dz} & c_{e2,\phi} \end{bmatrix}.
\end{equation}
The additive measurement noise $\mathbf{v}(t)$ is assumed white with covariance
\begin{equation}\label{eq:meas_noise_cov}
  \mathbb{E}[\mathbf{v}(t)\mathbf{v}^\mathrm{T}(s)] = \mathbf{R}\,\delta(t-s).
\end{equation}

The measurement-noise covariance matrix $\mathbf{R}$ is fully user-configurable and may include off-diagonal terms to represent correlations between measurement channels. In practical applications, $\mathbf{R}$ is typically chosen as a symmetric positive-definite covariance matrix. 

For well-decoupled systems, the user can choose to consider only a subset of degrees of freedom; the configuration algorithm then extracts the corresponding rows and columns from the state-space and covariance matrices to configure the LQG for a reduced system without requiring a redesign of the FPGA implementation.

\subsubsection{Discretization}\label{subsubsec:lqg_discretization}
For implementation on the FPGA, the system described by equations \eqref{eq:plant_ss_vec}--\eqref{eq:meas_noise_cov} has to be discretized at the sampling rate of $T_s = \SI{64}{\nano\second}$, yielding the discrete-time model
\begin{subequations}
\begin{equation}
\boldsymbol{\xi}[n+1] = \mathbf{A}_d\boldsymbol{\xi}[n] + \mathbf{B}_d\mathbf{u}[n] + \mathbf{w}_d[n],
\end{equation}
where $\boldsymbol{\xi}[n] = \boldsymbol{\xi}(t_0 +nT_s)$ denotes the sampled state vector and $\mathbf{w}_d[n]$ is a zero-mean discrete-time process-noise sequence with covariance

\begin{equation}
    \mathbb{E}[\mathbf{w}_d[n]\mathbf{w}_d^\mathrm{T}[m]] = \mathbf{Q}_d \delta_{nm}.
\end{equation}

We use the standard Van Loan method \cite{franklinDigitalControlDynamic1998,vanloanComputingIntegralsInvolving1978} to obtain an exact discretization of the linear dynamics and the corresponding process-noise covariance. The resulting discrete-time system matrices are
\begin{align}
\mathbf{A}_d &= e^{{\mathbf{A}} T_s}, \\
\mathbf{B}_d &= {\mathbf{A}}^{-1} (\mathbf{A}_d - \mathbf{I}) {\mathbf{B}}, \\
\mathbf{Q}_d &= \int_{0}^{T_s} e^{{\mathbf{A}}\tau} {\mathbf{G}}\mathbf{Q}\mathbf{G}^\mathrm{T}\left( e^{{\mathbf{A}}\tau} \right)^\mathrm{T} \, \mathrm{d}\tau
\end{align}
Since $\mathbf{R}$ denotes the continuous-time measurement-noise covariance density, discretization yields $\mathbf{R}_d = \mathbf{R}/{T_s}$.
\end{subequations}

\subsubsection{State estimation}
The optimal state estimator for the discrete-time linear system is the discrete-time Kalman filter \cite{brownIntroductionRandomSignals2012}. Writing the estimate of the state $\boldsymbol{\xi}$ at time instance $n$ as $\hat{\boldsymbol{\xi}}[n]$, the filter dynamics are given by
\begin{align}\label{eq:kalman}
  \hat{\boldsymbol{\xi}}[n+1] &= \mathbf{A}_d\hat{\boldsymbol{\xi}}[n]
      + \mathbf{B}_d\mathbf{u}[n]
      + \mathbf{L}_d\left(\,\boldsymbol{\chi}[n]-\mathbf{C}\hat{\boldsymbol{\xi}}[n]\,\right)
\end{align}
where the Kalman gain matrix $\mathbf{L}_d$ is obtained from the discrete-time algebraic Riccati equation as
\begin{gather}\label{eq:riccati_obs}
  \hat{\mathbf{P}}_d = \mathbf{A}_d\hat{\mathbf{P}}_d\mathbf{A}_d^\mathrm{T} - \mathbf{A}_d\hat{\mathbf{P}}_d\mathbf{C}^\mathrm{T}(\mathbf{C}\hat{\mathbf{P}}_d\mathbf{C}^\mathrm{T} + \mathbf{R}_d)^{-1}\mathbf{C}\hat{\mathbf{P}}_d\mathbf{A}_d^\mathrm{T} + \mathbf{Q}_d, \\
  \mathbf{L}_d = \mathbf{A}_d\hat{\mathbf{P}}_d\mathbf{C}^\mathrm{T}    (\mathbf{C}\hat{\mathbf{P}}_d\mathbf{C}^\mathrm{T} + \mathbf{R}_d)^{-1}.
\end{gather}

The Kalman filter reconstructs the full system state from the measured detector signals, providing estimates of positions, velocities, and the drift state required for feedback control. In typical optomechanical experiments, only position-related quantities are directly measured, making state estimation essential for reconstructing the full mechanical state required by the LQR feedback law.

\subsubsection{Linear quadratic regulator}
Control actions are obtained by solving a discrete-time linear quadratic regulator (LQR) problem \cite{ogataModernControlEngineering2010} over an infinite-time horizon. For a quadratic cost function
\begin{equation}\label{eq:lqr_cost}
  J = \lim_{N\rightarrow\infty} \frac{1}{N}\mathbb{E}\left[
    \sum_{n=0}^{N-1} \boldsymbol{\xi}^\mathrm{T}[n]\mathbf{Q}_{\mathrm{LQR}}
    \boldsymbol{\xi}[n] + \mathbf{u}^\mathrm{T}[n]\mathbf{R}_{\mathrm{LQR}}
    \mathbf{u}[n] \right],
\end{equation}
with state-weighting matrix $\mathbf{Q}_{\mathrm{LQR}}$ and feedback-action weighting matrix $\mathbf{R}_{\mathrm{LQR}}$, the resulting time-independent optimal full-state feedback law is
\begin{equation}\label{eq:lqr_law}
  \mathbf{u}[n] = -\mathbf{K}_d\,\boldsymbol{\xi}[n]
\end{equation}
where the feedback gain $\mathbf{K}_d$ is computed from the solution $\mathbf{P}_\xi$ of the discrete algebraic Riccati equation
\begin{gather}\label{eq:riccati_ctrl}
  \mathbf{P}_\xi = \mathbf{A}_d^\mathrm{T}\mathbf{P}_\xi\mathbf{A}_d - \mathbf{A}_d^\mathrm{T}\mathbf{P}_\xi\mathbf{B}_d(\mathbf{B}_d^\mathrm{T}\mathbf{P}_\xi\mathbf{B}_d + \mathbf{R}_{\mathrm{LQR}})^{-1}\mathbf{B}_d^\mathrm{T}\mathbf{P}_\xi\mathbf{A}_d + \mathbf{Q}_{\mathrm{LQR}}, \\
  \mathbf{K}_d = (\mathbf{B}_d^\mathrm{T}\mathbf{P}_\xi\mathbf{B}_d + \mathbf{R}_{\mathrm{LQR}})^{-1}\mathbf{B}_d^\mathrm{T}\mathbf{P}_\xi\mathbf{A}_d.
\end{gather}

The matrix $\mathbf{Q}_\mathrm{LQR}$ weighs the states proportionally to their mechanical energy via \linebreak $E = \frac{1}{2}\boldsymbol{\xi}^\mathrm{T}\mathbf{Q}_\mathrm{LQR}\boldsymbol{\xi}$, and is defined as
\begin{equation}
  \mathbf{Q}_\mathrm{LQR} = \mathbf{Q}_\mathrm{usr}\odot\left( \mathbf{q}_\mathrm{LQR}\mathbf{q}_\mathrm{LQR}^\mathrm{T}\right),
  \quad \mathbf{q}_\mathrm{LQR} = \sqrt{\frac{m}{2}}\begin{bmatrix}\left|\Omega_x \right| & 1 & \left|\Omega_y\right| & 1 & \left|\Omega_z\right| & 1 & 1 \end{bmatrix}^\mathrm{T}.
\end{equation}
It should be noted that the product $\mathbf{q}_\mathrm{LQR}\mathbf{q}_\mathrm{LQR}^\mathrm{T}$ produces a full matrix including off-diagonal cross terms. The matrix $\mathbf{Q}_\mathrm{usr}$ is applied to this full matrix with the element-wise product $\odot$ and is set to the identity matrix $\mathbb{I}_7$ by default, which recovers the standard mechanical energy-weighted diagonal LQR cost function. Reconfiguring the matrix $\mathbf{Q}_\mathrm{usr}$ allows the user to adjust the weighting of the individual states, for example, to prioritize cooling of a specific mode or to introduce off-diagonal terms to attenuate relative motion between modes.

This choice is particularly well suited for feedback-cooling applications, where the primary objective is the reduction of the mechanical energy stored in the oscillatory modes rather than the minimization of individual state components. Consequently, meaningful default LQR weights can be derived directly from the physical model, reducing the amount of manual controller tuning required by the user.

The matrix $\mathbf{R}_\mathrm{LQR}$ is fully user-configurable. Larger entries of $\mathbf{R}_\mathrm{LQR}$ penalize control effort more strongly, resulting in more conservative feedback actions, whereas smaller values permit more aggressive control.

\subsection{Implementation}\label{subsec:lqg_implementation}
This section provides an overview of implementation-specific aspects of the LQG controller described in the previous section. Its discrete-time input-output behavior is assembled from equations \eqref{eq:kalman} and \eqref{eq:lqr_law}, which yield
\begin{align}
  \hat{\boldsymbol{\xi}}[n+1] &= \left(\mathbf{A}_d-\mathbf{L}_d\mathbf{C}\right)\hat{\boldsymbol{\xi}}[n] + \mathbf{B}_d\mathbf{u}[n] + \mathbf{L}_d\boldsymbol{\chi}[n] \label{eq:lqg_eq1} \\
  \mathbf{u}[n] &= -\mathbf{K}_d\hat{\boldsymbol{\xi}}[n] \label{eq:lqg_eq2}.
\end{align}

\subsubsection{Fixed-point arithmetic}\label{subsubsec:fixpt_arith}
The SoC FPGA mounted on the Red Pitaya STEMLab Z7020 primarily supports fixed-point multiplication of 18- and 25-bit words. To balance the trade-off between resource consumption and precision loss due to quantization, we implemented barrel shifters for each multiplication, decomposing the operation into a product of the residual and a bit shift. The selected word lengths match the native 25×18 multiplier structure of the Xilinx DSP48E1 slices, thereby avoiding operation splitting and minimizing resource consumption and processing latency.

For the matrix product $\mathbf{C} = \mathbf{A}\mathbf{B}$, with $\mathbf{A} \in \mathbb{R}^{M \times K}$, $\mathbf{B} \in \mathbb{R}^{K \times N}$, and $\mathbf{C} \in \mathbb{R}^{M \times N}$, each element of the left matrix is encoded as a \textit{shift-float} number
\begin{equation}
    \tilde{a}_{ik} = a_{ik} 2^{-s_{ik}},
\end{equation}
where $a_{ik}$ is a fixed-point residue and $s_{ik} \in \mathbb{Z}$ is a
per-element unsigned integer exponent. Each element of the matrix $\mathbf{C}$ is obtained in
real-number terms from
\begin{equation}
    c_{ij} = \sum_{k=0}^{K-1} \tilde{a}_{ik} b_{kj}
           = \sum_{k=0}^{K-1} \left(a_{ik} b_{kj}\right)2^{-s_{ik}}.
\end{equation}
This approach combines some of the dynamic-range advantages of floating-point arithmetic with substantially lower hardware cost. By assigning an individual scaling exponent to each matrix coefficient, the most significant bits of each value can be preserved across a wide dynamic range while maintaining efficient FPGA utilization. The fixed-point data types configured for components in equations \eqref{eq:lqg_eq1} and \eqref{eq:lqg_eq2} are listed in Table~\ref{tab:lqg_fix_dtypes}.

\begin{table}[h]
\centering
\begin{tabular}{|l|c|c|c|}
\hline
& Bit width & Fraction width & Shift bits \\
\hline
\hline
Internal states $\hat{\boldsymbol{\xi}}$ & 25 & 22 & -- \\
\hline
Feedback signal $\mathbf{u}$ & 14 & 13 & -- \\
\hline
Detection signal $\boldsymbol{\chi}$ & 14 & 13 & -- \\
\hline
Dynamic matrix $\left(\mathbf{A}_d-\mathbf{L}_d\mathbf{C}\right)$ & 18 & 14 & 5 \\
\hline
Input matrix $\mathbf{B}_d$ & 18 & 14 & 5 \\
\hline
Kalman gain $\mathbf{L}_d$ & 18 & 14 & 5 \\
\hline
LQR gain $\mathbf{K}_d$ & 18 & 14 & 6 \\
\hline
\end{tabular}
\caption{Fixed-point data types for LQG components}
\label{tab:lqg_fix_dtypes}
\end{table}

\subsubsection{ADC and DAC calibration}
Accurate calibration of the input and output channels is required to ensure consistency between the physical model used for controller synthesis and the voltages processed by the FPGA implementation. To convert between raw digital values and physical voltage measurements, calibration parameters are applied to the input (ADC) and output (DAC) signals.

The \textbf{calibrated detection signal} $\boldsymbol{\chi}(k)$ (from the analog-to-digital converter) is computed as
\begin{equation}
  \boldsymbol{\chi}(k) = \left(\boldsymbol{\chi}_{\mathrm{raw}}(k) + \mathbf{o}_{\mathrm{in}}\right) \odot \mathbf{g}_{\mathrm{in}},
\end{equation}
where $\boldsymbol{\chi}_{\mathrm{raw}}(k)$ are the raw ADC readings, $\mathbf{o}_{\mathrm{in}}$ is the input offset vector, and $\mathbf{g}_{\mathrm{in}}$ is the input gain correction factor. The $\odot$ operator again denotes element-wise multiplication.

The \textbf{calibrated output signal} $\mathbf{u}_{\mathrm{out}}$ (to the digital-to-analog converter) is computed as
\begin{equation}
  \mathbf{u}_{\mathrm{out}}(k) = \mathbf{u}(k) \odot \mathbf{g}_{\mathrm{out}} + \mathbf{o}_{\mathrm{out}},
\end{equation}
where $\mathbf{u}(k)$ is the feedback control law output from equation~\eqref{eq:lqg_eq2}, $\mathbf{g}_{\mathrm{out}}$ is the output gain correction factor, and $\mathbf{o}_{\mathrm{out}}$ is the output offset vector added after scaling.

These calibration parameters compensate offset and gain mismatches of the analog front-end, ensuring that the FPGA operates on accurately scaled physical signals and that the generated control voltages correspond to the intended actuation levels.

\subsubsection{State-space transformation}

To improve numerical stability and accommodate the constraints of fixed-point arithmetic on the FPGA, the controller supports a sequence of regular state transformations. A major challenge in FPGA-based implementations is the large variation in magnitude between physical states, measurements, and controller gains. In particular, displacements, velocities, and measured voltages may differ by several orders of magnitude. Without additional scaling, this can lead to inefficient use of the available fixed-point dynamic range and increased quantization errors. To address this issue, the controller applies state-space transformations that improve numerical conditioning while preserving the input-output behavior of the system.

A regular state transformation $\bar{\boldsymbol{\xi}} = \mathbf{T}\boldsymbol{\xi}$ preserves the input-output behavior of the system while transforming the state-space matrices according to
\begin{equation}\label{eq:state_space_omega_transform}\
\bar{\mathbf{A}} = \mathbf{T}\mathbf{A}\mathbf{T}^{-1},\
\bar{\mathbf{B}} = \mathbf{T}\mathbf{B},\
\bar{\mathbf{C}} = \mathbf{C}\mathbf{T}^{-1},\
\bar{\mathbf{K}}_d = \mathbf{K}_d\mathbf{T}^{-1}.
\end{equation}
The controller supports several transformation variants configurable in the Graphical User Interface (GUI) to improve numerical stability and optimize the dynamic range of the fixed-point representation, as detailed in the following sections.

\subsubsection*{Physical state normalization}
This transformation is motivated by the fact that the state vector $\boldsymbol{\xi}$ contains both displacements and velocities, which can have vastly different numerical ranges, and additionally, the measurement matrix $\mathbf{C}$ can have entries with very large magnitudes, necessary to convert physical displacements, typically in the nanometer range, into measurable voltages.

The first transformation equalizes the numerical scale of displacements and velocities while simultaneously reducing the magnitude gap between physical states and measured voltages. To that end, the state-space matrices are transformed to first rebalance the magnitudes of the internal states and normalize the measurement matrix using its largest absolute entry $c_{e,\mathrm{max}}$. The diagonal normalization transformation matrix is defined as
\begin{equation}
\mathbf{T}_\mathrm{n} = \mathrm{diag}\left( c_{e,\mathrm{max}}\cdot \begin{bmatrix}
1 & 1/\left|\Omega_x\right| & 1 & 1/\left|\Omega_y\right| & 1 & 1/\left|\Omega_z\right| & 1
\end{bmatrix} \right).
\end{equation}

\subsubsection*{Numerical state-space transformation}
For robust operation on fixed-point hardware, an additional transformation rescales the states to exploit the full dynamic range of the fixed-point representation. Unlike heuristic scaling approaches, the proposed balancing procedure derives the scaling factors directly from the predicted closed-loop covariance, enabling automated normalization of the internal state magnitudes without manual tuning.

The system is scaled by matching the estimated closed-loop state covariance obtained from the discrete-time Lyapunov equation \cite{lewisOptimalRobustEstimation2017}
\begin{equation}
\mathbf{0} = \left(\mathbf{A}_d-\mathbf{B}_d\mathbf{K}_d\right)\mathbf{P}_{\hat{\xi}}\left(\mathbf{A}_d-\mathbf{B}_{d}\mathbf{K}_d\right)^\mathrm{T}  - \mathbf{P}_{\hat{\xi}} +  \mathbf{L}_d\mathbf{R}_d\mathbf{L}_d^\mathrm{T},
\end{equation}
to a user-defined target standard deviation $\bar{{\sigma}}_\xi$, which defines the diagonal target state standard-deviation matrix as
\begin{equation}
\bar{\boldsymbol{\Sigma}}_\xi = \mathbb{I} \cdot \bar{\sigma}_\xi.
\end{equation}
With $\mathbf{S}$ obtained from Cholesky decomposition $\mathbf{P}_{\hat{\xi}} = \mathbf{S}^\mathrm{T}\mathbf{S}$, the balancing transformation matrix is simply calculated as
\begin{equation}
\mathbf{T}_{\mathrm{bal}} = \mathrm{diag}(\mathbf{S}\bar{\boldsymbol{\Sigma}}_\xi^{-1}).
\end{equation}

Note that while the $\mathbf{S}\bar{\boldsymbol{\Sigma}}_\xi^{-1}$ matrix is generally not diagonal, only its diagonal elements are retained. This preserves the physical interpretation of the individual states while simplifying data handling and controller configuration.

To prevent overflow in fixed-point arithmetic, the transformation is further clamped to ensure that the magnitudes of the input matrix $\mathbf{B}_d$, the Kalman gain $\mathbf{L}_d$, and the feedback gain $\mathbf{K}_d$ remain within the available numerical range of the fixed-point representation, namely $\pm1.0$ and $\pm31.0$, respectively, leading to the input-output transformation matrix
\begin{equation}
\mathbf{T}_{\mathrm{InOut}} =\begin{cases}
  \mathbb{I} \cdot \max\left(\begin{bmatrix}\left|\mathbf{T}_{\mathrm{bal}}\mathbf{B}_d\right| & \left|\mathbf{T}_{\mathrm{bal}}\mathbf{L}_d\right| \end{bmatrix} \right),& 
  \text{if }\max\left(\begin{bmatrix}\left|\mathbf{T}_{\mathrm{bal}}\mathbf{B}_d\right| & \left|\mathbf{T}_{\mathrm{bal}}\mathbf{L}_d\right| \end{bmatrix} \right)>1 \\
  \mathbb{I} \cdot 1/\max\left(\left|\mathbf{K}_d\mathbf{T}_{\mathrm{bal}}^{-1}\right|\right),& 
  \text{if }{\max\left(\left|\mathbf{K}_d\mathbf{T}_{\mathrm{bal}}^{-1}\right|\right)}>31 \\
  \mathbb{I}, &\text{otherwise},
\end{cases} 
\end{equation}
where $\max(|\mathbf{M}|)$ denotes the largest absolute matrix element of $\mathbf{M}$. The final numerical transformation is obtained as
\begin{equation}
\mathbf{T}_{\mathrm{num}} = \mathbf{T}_{\mathrm{bal}}\mathbf{T}_{\mathrm{InOut}}
\end{equation}
and applied to the state space system analogously to \eqref{eq:state_space_omega_transform}. The resulting matrices are used for the fixed-point implementation, and the inverse transformation is applied to the internal states before outputting them to the GUI, ensuring that the user interacts with physical values that can be optionally scaled by the physical state normalization transformation $\mathbf{T}_\mathrm{n}$ for improved readability.

\subsubsection{Parameter configuration}
We leverage the integration of the FPGA and processor offered by the SoC FPGA (Xilinx Zynq Z7020), enabling live controller configuration, data acquisition, and parameter updates without requiring FPGA recompilation.

A Python REST API server runs on the processor, and communication is based on HTTP POST requests. For the configuration of the LQG, the physical parameters set by the user are processed by the GUI, the system is discretized, and the feedback and Kalman gains are calculated automatically. The resulting matrices are then decomposed into their shift and residual components (see Section \ref{subsubsec:fixpt_arith}) and packaged for deployment to the FPGA. To ensure deterministic controller operation, all matrices are first staged in intermediate registers and subsequently updated simultaneously. The GUI further supports storing two complete parameter sets, enabling rapid switching between different controller configurations.

The implementation also allows recording all input and output signals, as well as all internal states, at full precision. The individual channels are latched at a single time instance (frame) at a user-configured multiple of the base sampling time of $\SI{64}{\nano\second}$. The user can also configure the number of frames to be recorded and select which channels to record to avoid saving unnecessary data, such as unused degrees of freedom. The recorded data is stored in onboard memory until it is requested by the GUI, which transfers the requested amount of data for plotting or export. The data capacity depends on the amount reserved by the user (as detailed in Section \ref{subsubsec:lqg_hw_setup}).

\input{pages/lqg/lqg_setup}

%% file: pages/lqg/lqg_setup.tex
\subsection{Setup}\label{subsec:lqg_setup}
The prerequisites for running the LQG are:
\begin{itemize}
\item RedPitaya STEMlab 125-14 Z7020 (Gen 1 or 2) with OS 2.07-43 or newer.
\item MATLAB 2024a or newer, or MATLAB Runtime R2024a when using the compiled GUI.
\item Compatible FPGA bitstream \texttt{lqg\_fp7\_vX.Y.Z.bin}.
\item \textbf{Optional:} MATLAB toolboxes and add-ons required to run the GUI source code: Control System Toolbox, Image Processing Toolbox, Optimization Toolbox, Signal Processing Toolbox, and Advanced Logger for MATLAB 2.0.2.
\item \textbf{Optional:} Internet connection for the Red Pitaya.
\item \textbf{Optional:} Oscilloscope and signal generator for IO calibration.
\end{itemize}

Additionally, if you wish to use the data recording feature of the LQG, make sure you modify the device tree according to the following section.

\subsubsection{Red Pitaya memory reservation}
To support the data recording feature of the LQG GUI, capacity must be reserved in onboard DDR memory; we recommend at least 256~MB. The workflow differs between Gen 1 and Gen 2 Red Pitaya models, as described in the following sections. Make sure to configure the GUI according to the amount of reserved memory.

\paragraph{Gen1 Red Pitaya Z20-125.}
\begin{enumerate}
    \item Unmount the SD card from the Red Pitaya (RP) and mount it on a PC.
    \item Navigate to the boot partition and open the file \texttt{dts/z20\_125/dtraw.dts}.
    \item Find the following text in the \texttt{memory} section:
    \begin{minted}{dts}
    ...
    memory {
        device_type = "memory";
        reg = <0x00 0x20000000>;
    };
    ...
    \end{minted}
    \item Edit the \texttt{reg} parameter as:
    \begin{minted}{dts}
        reg = <0x00 0x10000000>;
    \end{minted}
    \item Save the changes and compile the device tree using: \\
    \texttt{dtc -O dtb -o devicetree.dtb dtraw.dts}
    \item Save the changes, eject the SD card, and mount it back in the RP.
    \item After booting the RP up, verify the memory configuration using \texttt{head /proc/meminfo}.
\end{enumerate}

\paragraph{Gen2 Red Pitaya Z20-125.}
\begin{enumerate}
    \item Switch the RP to the 1~GB boot mode via the web GUI: click the information button in the top-left corner, open \textbf{System settings}, set \textbf{BOOT mode} to \textbf{1GB RAM}, and reboot the RP.
    \item Remount the boot partition as read/write using \\ \texttt{mount -o remount,rw /boot}.
    \item Open \texttt{/boot/dts/z20\_125\_v2/dtraw.dts} in a text editor.
    \item In the \texttt{reserved-memory} node at the end of the file, add the following configuration to reserve 512~MB:
    \begin{minted}{dts}
    dma_reserved@20000000 {
       compatible = "shared-dma-pool";
       no-map;
       reg = <0x20000000 0x20000000>;
    };
    \end{minted}
    \item Save the file, remount \texttt{/boot} as read-only using \texttt{mount -o remount,ro /boot}, and reboot the RP.
    \item Verify the change by checking that \texttt{ls /proc/device-tree/reserved-memory} lists \texttt{dma\_reserved@20000000}, \texttt{cat /proc/cmdline} does not contain a setting such as \texttt{mem=512MB}, and \texttt{cat /proc/iomem} lists \texttt{System RAM} only up to address \texttt{0x1fffffff}.
\end{enumerate}

\subsubsection{Hardware}
\label{subsubsec:lqg_hw_setup}

\begin{enumerate}
    \item Ensure the RP is accessible via SSH and SCP. We recommend using WinSCP and PuTTY for Windows.
    \item Make sure the following packages are installed in the OS of the RP:
        \begin{itemize}
            \item \texttt{python3-yaml}
            \item \texttt{python3-bitarray}
            \item \texttt{python3-numpy}
            \item \texttt{python3-flask}
        \end{itemize}
        If the RP has internet access, install them using \\ \texttt{apt-get install python3-yaml python3-bitarray \\ python3-numpy python3-flask}

    \item Copy the distribution package \texttt{package/lqg\_fp7\_package.tar.gz} to the RP using SCP.

    \item Extract the package on the RP using: \\
        \texttt{tar -xf lqg\_fp7\_package.tar.gz}

    \item Run \texttt{./start.sh} to initialize the FPGA bitstream and start the server.
\end{enumerate}

\subsubsection{Host PC setup}
\label{subsubsec:lqg_sw_setup}
\begin{enumerate}
    \item Clone the \texttt{LQG/user-interface} repository to your host PC.
    \item Navigate to the \texttt{user-interface} folder.
    \item If you only want to run the compiled application without a MATLAB license, install it together with MATLAB Runtime via \texttt{/standalone/for\_redistribution/LQG-FP7\_Installer\_web.exe}.
    \item If MATLAB Runtime R2024a is already installed, you can start the compiled application directly from \texttt{/standalone/for\_redistribution\_only/LQG\_FP7\_GUI.exe}.
    \item Alternatively, open the source code \texttt{LQG\_FP7\_GUI.mlapp} in MATLAB and start the GUI from there.
\end{enumerate}

To get started with the GUI, refer to the example in Section \ref{sec:example}. For technical documentation of the GUI, refer to \texttt{LQG\_FP7/user-interface/docs} (also accessible from within the \textbf{User guide} tab in the GUI).

%% file: pages/simulator_sections.tex
\section{HIL Simulator}\label{sec:simulator}
The HIL simulator is intended to validate implementations of real-time controllers in a safe and reproducible environment that closely resembles the behavior of the target experiment. To this end, we developed a general architecture capable of simulating three-degree-of-freedom systems with nonlinear dynamics and nonlinear input-output relationships, thereby enabling controller validation beyond the linear models used for controller synthesis.

First, the mathematical model is introduced in generalized form and then specialized to levitated nanoparticles. Next, the implementation is presented, including the overall architecture and timing constraints, the in- and output matrices, nonlinear function realization, state-slice integration, and noise generation. Finally, the fixed-point datatypes and scaling choices used in the FPGA implementation are summarized.

\subsection{Mathematical Model}\label{subsec:sim_model}
The presented HIL simulator is designed to solve stochastic differential equations (SDEs), where the state evolution is governed by both deterministic dynamics and stochastic noise processes \cite{jacobs:2010}. To ensure architectural flexibility encompassing a wide array of experiments, the simulator implements a general system model which includes specific physical systems like levitated nanoparticles.

We define a second-order generalized stochastic model that accommodates nonlinear damping, potential fields, and input/output nonlinearities per $N_\textrm{st}=3$ degrees of freedom (DOF), with $N_\textrm{in}=2$ inputs and $N_\textrm{out}=2$ outputs. The model is given by
\begin{subequations}\label{eq:generalized_system_model}
\begin{align}
    \ddot{q}_j(t) &= f_j(\dot{q}_j(t)) + g_j(q_j(t)) + c_j z_j(t), \quad j = 1, \dots, N_\textrm{st}, \\
    z_j(t) &= b_j \xi_j(t) + \sum_{i=1}^{N_\textrm{in}} a_{j,i} \alpha_i(u_i(t)),
\end{align}
where $q_j$ represents the state variables, $f_j(\dot{q}_j(t))$ the nonlinear damping function, $g_j(q_j(t))$ the nonlinear potential function, $c_j$ the coupling coefficient of the $z_j$ exogenous input, assembled from the control inputs $u_i$ fed through an input nonlinearity $\alpha_i$ and weighted by input coefficients $a_{j,i}$ for each state. The stochastic component is represented by the noise term $b_j \xi_j(t)$, providing an independent unitary Gaussian white noise source for each state, where $b_j$ is the noise scaling factor. 

The output $y_k$ is formed by summing contributions from each degree of freedom, where the state variables are passed through an output nonlinearity $\beta_{j,k}(q_j(t) ; \dot{q}_j(t))$ that can select either the position $q_j$ or velocity $\dot{q}_j$ and are weighted by coefficients $d_{k,j}$, following the form
\begin{equation}
    y_k(t) = \sum_{j=1}^{N_\textrm{st}} d_{k,j} \beta_{j,k}(q_j(t) ; \dot{q}_j(t)) \quad k = 1, \dots, N_\textrm{out}.
\end{equation}
\end{subequations}

Many physical systems exhibit coupled nonlinear dynamics and more intricate structures. However, implementing fully coupled multivariate nonlinearities would substantially increase LUT, memory, and routing costs and reduce the timing margin on the target FPGA. For this reason, the presented architecture uses three independent state slices and separable nonlinearities in both the state and input/output paths. This design choice limits model generality but preserves real-time performance and reconfigurability, while still covering a broad class of systems with nonlinear yet approximately separable dynamics without hardware-level redesign.

One system belonging to this class is the center-of-mass motion of a nanoparticle levitated in an optical trap. Such systems are governed by the Langevin equation, which balances viscous damping, nonlinear (usually restoring) forces from the optical potential $V(x)$ \cite{haradaRadiationForcesDielectric1996}, and stochastic Brownian motion \cite{millenOptomechanicsLevitatedParticles2020a}. Including an external electrostatic feedback force $F_u(t)$, the equation of motion in one dimension is \cite{gieselerNonequilibriumSteadyState2015}
\begin{equation}
    \label{eq:particle_full_equation}
    m \dv[2]{x}{t} = - \gamma \dv{x}{t} - \nabla V(x) + F_u(t) + \sqrt{2 \gamma k_B T} \xi(t).
\end{equation}
Here, $m$ is the particle mass and $\gamma$ is the damping coefficient, while $k_B$, $T$, and $\xi(t)$ denote the Boltzmann constant, the environmental temperature, and a unit Gaussian white-noise process, respectively. This physical model maps directly to \eqref{eq:generalized_system_model} by setting $f(\dot{q})$ to linear damping, $g(q)$ to the potential gradient, and $b = \sqrt{2 \gamma k_B T}$ to the thermal noise.

\subsection{Implementation}
Levitated optomechanical systems present demanding parameters with quality factors up to $10^8$ at low pressures \cite{millenOptomechanicsLevitatedParticles2020a} and high natural frequencies. Reaching hundreds of kHz with minimal damping requires a numerical integrator to solve the differential equation at high sampling rates and numerical stability to avoid divergence in real-time execution. Additionally, the input-output latency of the HIL simulator must be kept low to allow testing of feedback control algorithms. The simulator architecture is therefore designed to balance numerical accuracy, low latency, and FPGA resource efficiency while maintaining real-time execution at the target sampling rate.

\subsubsection{Architecture overview}
\label{sec:architecture_overview}

The core of the simulator's architecture comprises two analog inputs and two analog outputs, each with its own ADC and DAC, as well as the main processing core, implemented as a digital system processing (DSP) algorithm on an FPGA. The architecture is designed to minimize input-output latency while maintaining sufficient numerical accuracy for hardware-in-the-loop validation of feedback controllers.

The simulator's DSP system solves the system model \eqref{eq:generalized_system_model} numerically in real time to obtain $q[n]$ and $y[n]$ every time step $n$. The architecture can be split up into the following parts:

\begin{itemize}
    \item \textbf{The noise generators}, which generate pseudorandom Gaussian white noise
    signals $\xi_j[n]$ required for simulating stochastic systems. For each of the $N_\textrm{st}$ exogenous inputs $z_j[n]$, there is a corresponding noise source $\xi_j[n]$.

    \item \textbf{The input mapping}, which maps the input signals $u_i[n]$ to the three exogenous inputs $z_j[n]$. Additionally, the corresponding noise signals $\xi_j[n]$ are added to the exogenous inputs.

    \item \textbf{The three state processing slices}, each of which computes one model state $q_j[n]$ and its derivative $\dot{q}_j[n]$ given the state exogenous inputs $z_j[n]$ by integrating the state equations in \eqref{eq:generalized_system_model} forward in time by one time step. The use of three independent state-processing slices follows directly from the separable model structure introduced in Section \ref{subsec:sim_model}, enabling efficient parallel implementation on the FPGA.

    \item \textbf{The output mapping}, which maps the state outputs $q_j[n]$ and
    $\dot{q}_j[n]$ to the simulator outputs $y_k[n]$.
    
\end{itemize}

\begin{figure}[H]
	\centering
	\def\svgscale{1.2}
    \captionsetup{width=0.9\linewidth,format=hang}
	\graphicspath{ {graphics/} }
	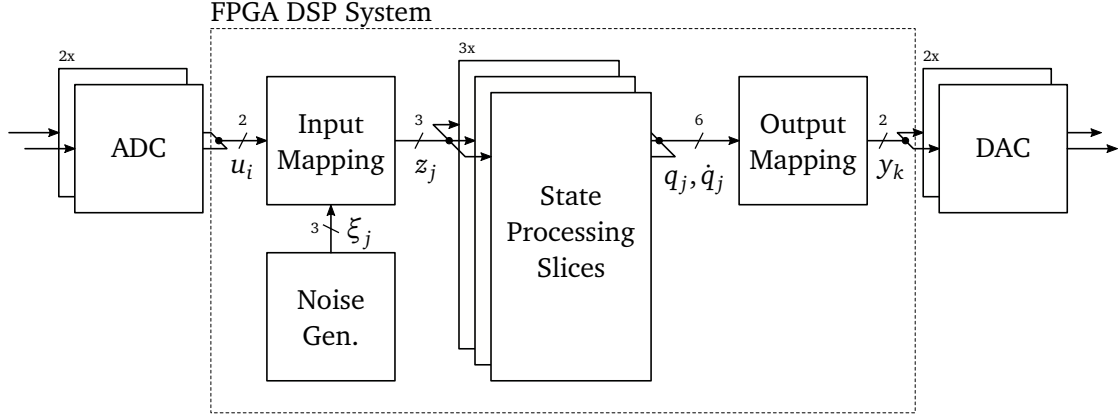
    \caption{Topological overview of the HIL simulator architecture.}
    \label{fig:architecture_overview}
\end{figure}

A topological overview of the architecture is shown in Figure \ref{fig:architecture_overview}. Due to FPGA resource limitations, the number of independent state-processing slices was set to $N_\textrm{st} = 3$, covering typically three translational degrees of freedom. The architecture is designed to be modular and scalable, allowing for future extensions to more states, inputs, or outputs if supported by the target hardware.

The system is discretized at a sampling rate $\num{6,944}\,\text{MS/s}$. The breakdown of delays across all processing sections is summarized in Table \ref{tab:processing_delay}. NLF denotes a configurable nonlinear function block.

\begin{table}[H]
    \centering
    \def\arraystretch{1.3}
    \captionsetup{width=0.9\linewidth,format=hang}
    \begin{tabular}{lcc}
        Section & $\num{4}\,\si{\nano\second}$ Clk. Cycles & Time \\
        \hline
        \hline
        Apply ADC Calibration                       & $\num{4}$     & $\num{16}\,\si{\nano\second}$     \\
        input mapping - NLF (worst-case)             & $\num{12}$    & $\num{48}\,\si{\nano\second}$     \\
        input mapping - Linear                       & $\num{6}$     & $\num{24}\,\si{\nano\second}$     \\
        Processing Slices                           & $\num{36}$    & $\num{144}\,\si{\nano\second}$    \\
        output mapping - NLF (worst-case)            & $\num{12}$    & $\num{48}\,\si{\nano\second}$     \\
        output mapping - Linear                      & $\num{7}$     & $\num{28}\,\si{\nano\second}$     \\
        Apply DAC Calibration                       & $\num{4}$     & $\num{16}\,\si{\nano\second}$     \\
        \hline
        \textbf{Total Processing Delay} $t_\mathrm{pd,p}$    & $\num{81}$    & $\num{324}\,\si{\nano\second}$    \\
        \hline
        ADC+DAC conversion time $t_\mathrm{pd,conv}$         & $\num{116}$ & $\num{464}\,\si{\nano\second}$     \\
        \hline
        \textbf{Total worst-case input-output delay including processing} $t_\mathrm{pd,tot}$ & $\num{197}$ & $\num{788}\,\si{\nano\second}$ \\
         \hline
    \end{tabular}
    \caption{
        Decomposition of the total processing delay of the implemented algorithm and the resulting input-output delay including the ADC and DAC conversion time of the HIL Simulator.
    }
    \label{tab:processing_delay}
\end{table}

\subsubsection{In- and output mapping}
\label{sec:2_input_and_output_matrices}

\begin{figure}
	\centering
	\def\svgscale{1.2}
    \captionsetup{width=0.9\linewidth,format=hang}
	\graphicspath{ {graphics/} }
	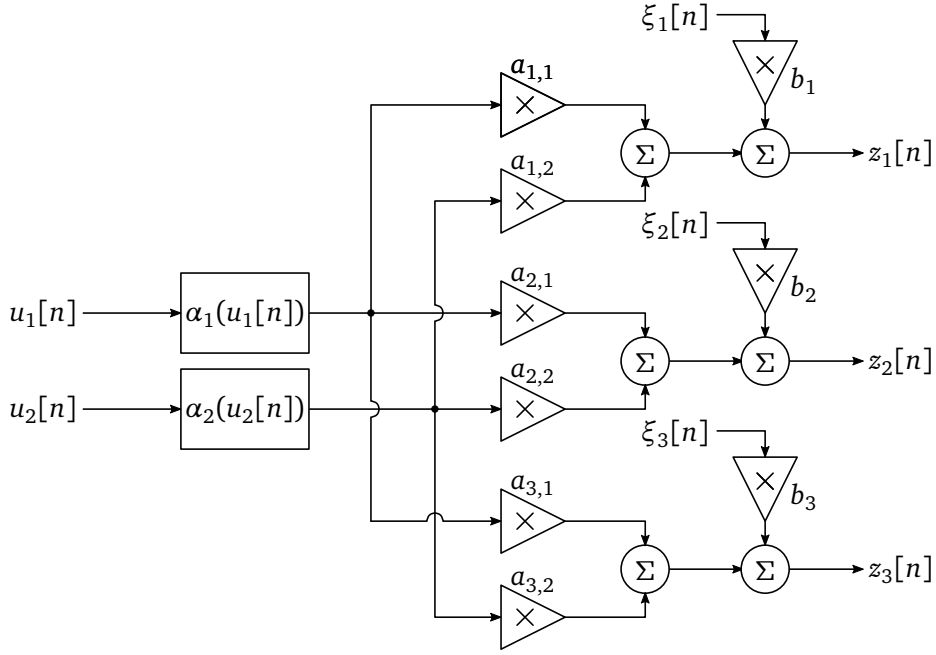
    \caption{
        Input mapping architecture with two simulator inputs $u_{i}[n]$ and
        three state inputs $z_{j}[n]$.
    }
    \label{fig:input_matrix_arch}
\end{figure}

\paragraph{Input mapping}
The input mapping maps the simulator inputs $u_{i}[n]$ to the intermediate signals $z_{j}[n]$, which ultimately enter as acceleration in the state-processing slices according to Equation~\eqref{eq:input_mapping} as 
\begin{equation}
    \label{eq:input_mapping}
    z_j[n] = b_j \xi_j[n] + \sum_{i=1}^{N_\textrm{in}} a_{j,i} \alpha_i(u_i[n]) \quad j = 1, \dots, N_\textrm{st}
\end{equation}
The inputs $u_{i}[n]$ pass through the nonlinear functions $\alpha_i$ and then enter the individual states with relative weights $a_{j,i}$, allowing to weigh the inputs independently for each state. Additionally, stochastic inputs $\xi_j$ scaled by $b_j$ are added to the signal. Each $\alpha_i(\cdot)$ acts on one input channel only (as shown in Fig.~\ref{fig:input_matrix_arch}); cross-channel mixing is implemented by the weights $a_{j,i}$ in the matrix stage.

Its implementation is shown in Figure~\ref{fig:input_matrix_arch}. To minimize propagation delay, the input mapping operates on sample data at the full clock rate of $250\,\si{\mega\hertz}$.

\paragraph{Output mapping}
The output mapping gathers the state variables $q_{j}[n]$ and $\dot{q}_{j}[n]$ from the slices and forms the simulator outputs $y_{k}[n]$ as specified in Equation~\eqref{eq:output_mapping}.
\begin{equation}
    \label{eq:output_mapping}
    y_k[n] = \sum_{j=1}^{N_\textrm{st}} d_{k,j} \beta_{j,k}(q_j[n] ; \dot{q}_j[n]) \quad k = 1, \dots, N_\textrm{out}.
\end{equation}
This produces an output voltage $y_k$ as a weighted combination of the nonlinear functions $\beta_{j,k}$, selecting either the displacement or the velocity of the DOFs, reflecting typical measurements in real systems.

\begin{figure}
	\centering
	\def\svgscale{1.2}
    \captionsetup{width=0.9\linewidth,format=hang}
	\graphicspath{ {graphics/} }
	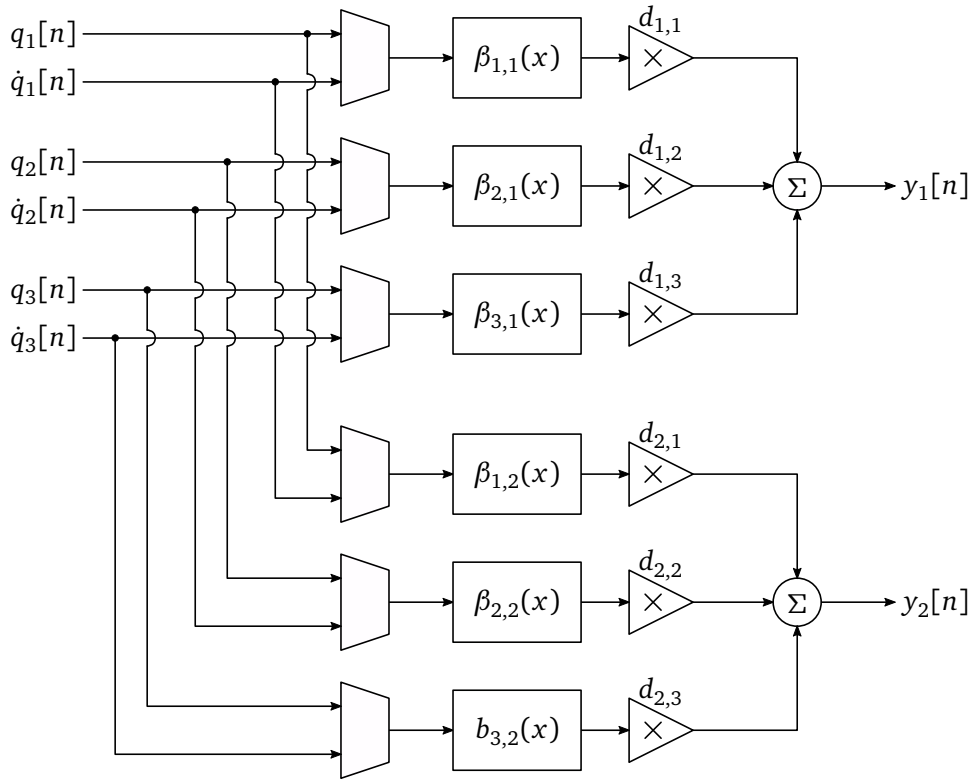
    \caption{
        Output mapping architecture with three state variables $q_{j}[n]$ and
        $\dot{q}_{j}[n]$ and two simulator outputs $y_{k}[n]$.
    }
    \label{fig:output_matrix_arch}
\end{figure}

Its implementation is shown in Figure~\ref{fig:output_matrix_arch}. Like the input mapping, it processes sample data at the full clock rate of $250\,\si{\mega\hertz}$ to reduce overall propagation delay.

\subsubsection{State processing slices}
\label{sec:2_state_slices_and_feedback_synchronization}

Each state processing slice implements the numerical integration of a single degree of freedom $q_j$ given by 
\begin{equation}
    \label{eq:state_slices}
    \ddot{q}_j[n] = f_j(\dot{q}_j[n]) + g_j(q_j[n]) + z_j[n], \quad j = 1, \dots, N_\textrm{st},
\end{equation}
where $f_j$ and $g_j$ are configurable nonlinearities, and $z_j$ denotes the exogenous input. Thus, the slice realizes the normalized, fixed-point form of the generalized model in Equation \eqref{eq:generalized_system_model}.

The acceleration in \eqref{eq:state_slices} is integrated by a combined Adams-Bashforth and Adams-Moulton scheme \cite{hairer:2006}
\begin{equation}
\label{eq:abam3}
\begin{aligned}
    \dot{q}[n+1] &= \dot{q}[n] + t_s \!\left( \tfrac{3}{2}\,\ddot{q}[n] - \tfrac{1}{2}\,\ddot{q}[n-1] \right), \\
    q[n+1] &= q[n] + \tfrac{t_s}{2}\!\left( \dot{q}[n+1] + \dot{q}[n] \right).
\end{aligned}
\end{equation}

\begin{figure}[H]
	\centering
	\def\svgscale{1.2}
    \captionsetup{width=0.9\linewidth,format=hang}
	\graphicspath{ {graphics/} }
	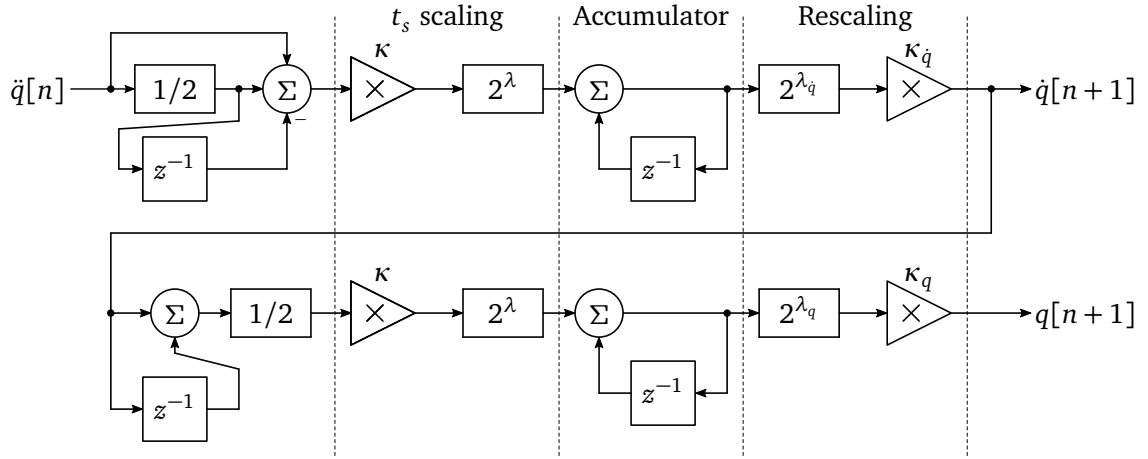
    \caption{
        Semi-implicit linear multistep integrator block with scaling implementing the discrete-time differential equation solver given by \eqref{eq:abam3}.
    }
    \label{fig:integrator_sim_lm2}
\end{figure}

This semi-implicit structure is used in each state-processing slice and corresponds to the integrator shown in Figure~\ref{fig:integrator_sim_lm2}. The explicit Adams--Bashforth step advances the velocity $\dot{q}$ to the next time step from the acceleration samples $\ddot{q}$, after which the implicit Adams--Moulton step updates the position $q$ using the newly available velocity sample. This combination improves numerical stability while retaining moderate implementation complexity.

Since the algorithm is implemented in fixed-point arithmetic, special care must be taken to accommodate the small sampling time ($t_s \approx 10^{-7}\,\mathrm{s}$). Hence, timestep factors are decomposed as $t_s = \kappa\,2^{\lambda}$, so products like $\dot{q}[n]\,t_s$ are computed as $(\dot{q}[n]\cdot\kappa) \ll \lambda$, with the arithmetic left-shift operator $\ll$. Bit-shift scalers efficiently implement a large dynamic range. Figure~\ref{fig:integrator_sim_lm2} shows the final integrator topology. The upper path implements the Adams–Bashforth velocity update, while the lower path implements the Adams–Moulton position update. The $\kappa$ and $2^{\lambda}$ blocks provide the fixed-point time-step scaling required for numerically robust accumulation and subsequent recovery of the physical state magnitude.

Stability and convergence properties of the chosen integrator are discussed in \cite{hairer:2006}. Since the Gaussian noise enters the model only additively, no separate evaluation of the integration scheme with respect to the noise term is necessary \cite{gardiner:2009}.

\subsubsection{Nonlinear Functions}
\label{sec:2_nonlinear_functions}
The nonlinear functions $f_j$, $g_j$, $\alpha_i$, and $\beta_{j,k}$ are implemented using lookup tables (LUTs) and linear interpolation between entries, providing a good trade-off between approximation accuracy and FPGA resource usage. LUT-based evaluation was selected because it enables arbitrary nonlinear functions to be implemented with deterministic latency and predictable FPGA resource usage, independent of the functional form. The function domain is partitioned into equally spaced intervals, and within each interval the function is approximated linearly, such as

\begin{equation}
 f(x) \approx f(x_i) + \frac{f(x_{i+1})-f(x_i)}{x_{i+1}-x_i}(x-x_i).
\end{equation}
The two neighboring LUT entries $f(x_i)$ and $f(x_{i+1})$ are fetched, the output is interpolated using the fractional remainder, and the result is then rescaled to maintain consistency.

Operationally, the input fixed-point number (e.g., the displacement $q_j$) is first scaled to the normalized interval $[-1,1]$. Due to the implementation, only zero-symmetric input intervals are supported. For a LUT with $2^N$ points, the $N$ most significant bits of the fixed-point input form the address of the nearest table entries, while the remaining lower bits serve as the interpolation fraction. In the present implementation, a 10-bit LUT address space, corresponding to $2^{10}$ entries, with a 16-bit data width was selected due to hardware resource constraints.  

The LUTs are realized as two-port RAM blocks in the FPGA fabric for flexibility. The stored function can be reconfigured at runtime, and the interpolation granularity can be adjusted to optimize accuracy versus resource usage \cite{lachowicz:2008}. 

\subsubsection{Noise generators}
\label{sec:2_noise_generator}
The simulator requires Gaussian-distributed white noise for simulating stochastic dynamics. This is generated using a Linear Feedback Shift Register (LFSR) for pseudorandom sequences \cite{singer:2022} and the Box-Muller transform for converting uniform to a Gaussian distribution \cite{lee:2006}.

Galois-type LFSRs operate on binary signals using XOR operations and can be efficiently implemented in hardware. A uniform-distributed pseudorandom sequence is generated at the slice sampling rate ($\sim$~7~MHz) using a Galois-type LFSR, with a 24-bit width. A 24-bit LFSR was selected as a compromise between sequence length, statistical quality, and FPGA resource consumption. To achieve the maximum sequence period for a given LFSR width, the feedback configuration must use a primitive polynomial \cite{koopman:checksums_cyclic_redundancy}, which is chosen as
\begin{equation}
x^{24} + x^{23} + x^{22} + x^{17} + 1.
\end{equation} 

To convert LFSR-generated uniform random numbers to Gaussian-distributed noise, the Box-Muller transform is applied \cite{box:1958}. This method takes two independent uniform random variables $u_1$ and $u_2 \in (0, 1)$ and produces two independent normal-distributed variables $x_1$, $x_2$ via \cite{scott:2011}

\begin{equation}
    \begin{aligned}
        x_1 &= \sqrt{-2 \ln(u_1)} \cos(2 \pi u_2) \\
        x_2 &= \sqrt{-2 \ln(u_1)} \sin(2 \pi u_2).
    \end{aligned}
\end{equation}

The two Box-Muller outputs are combined and rescaled to obtain the required zero-mean, unit-power Gaussian white-noise signal $\xi$, which is passed to the input mapping. The resulting noise sample is computed as
\begin{equation}
    \xi = c_\xi (x_1 + x_2),
\end{equation}
where $c_\xi$ is a scaling factor chosen such that $\xi$ has unit power. Since the simulator requires three independent noise inputs, three separate instances of this noise generator are implemented.

\subsubsection{Fixed-point datatypes}
\label{sec:2_fixed_point_datatypes}

The numerical performance of the DSP pipeline is determined by the chosen fixed‑point formats and their scaling. Because suitable scales depend on the specific plant model, they are computed by the configuration algorithm. To match the Xilinx ZYNQ DSP48E1 resources, which provide $25 \times 18$ multipliers \cite{amd:dsp48e1}, all internal state and signal paths use $\num{25}\,\si{\bit}$ words and all configurable constants use $\num{18}\,\si{\bit}$ words, avoiding cascaded multipliers and the associated timing penalty.

External I/O width is set by the converters: both ADC and DAC operate at $\num{14}\,\si{\bit}$. Two specialized formats are used where required: a $\num{47}\,\si{\bit}$ integration accumulator to maintain headroom during multistep updates, and $\num{16}\,\si{\bit}$ words for nonlinear function LUT entries. Together, these choices balance precision, dynamic range, resource usage, and clock performance; the fraction-point choices are summarized in Table~\ref{table:datatype_overview}.

\begin{table}[H]
    \centering
    \def\arraystretch{1.3}
    \captionsetup{width=0.9\linewidth,format=hang}
    \begin{tabular}{|l|c|c|}
        \hline
        \textbf{Datatype} & \textbf{Bit-Size} & \textbf{Decimal Point} \\
        \hline
        \hline
        In- and output signal datatype & $\num{14}\,\si{\bit}$ & $\num{12}\,\si{\bit}$ \\
        \hline
        General state and signal datatype & $\num{25}\,\si{\bit}$ & $\num{17}\,\si{\bit}$ \\
        \hline
        Datatype for configurable constants & $\num{18}\,\si{\bit}$ & $\num{10}\,\si{\bit}$ \\
        \hline
        Integration accumulator datatype & $\num{47}\,\si{\bit}$ & $\num{39}\,\si{\bit}$ \\
        \hline
        Nonlinear-Function-LUT word datatype & $\num{16}\,\si{\bit}$ & $\num{14}\,\si{\bit}$ \\
        \hline
    \end{tabular}
    \caption{
        Overview of all datatypes used in the HIL simulator implementation.
    }
    \label{table:datatype_overview}
\end{table}

\input{pages/simulator/model_config_setup.tex}

%% file: graphics/architecture_overview.pdf_tex
\begingroup%
  \makeatletter%
  \providecommand\color[2][]{%
    \errmessage{(Inkscape) Color is used for the text in Inkscape, but the package 'color.sty' is not loaded}%
    \renewcommand\color[2][]{}%
  }%
  \providecommand\transparent[1]{%
    \errmessage{(Inkscape) Transparency is used (non-zero) for the text in Inkscape, but the package 'transparent.sty' is not loaded}%
    \renewcommand\transparent[1]{}%
  }%
  \providecommand\rotatebox[2]{#2}%
  \newcommand*\fsize{\dimexpr\f@size pt\relax}%
  \newcommand*\lineheight[1]{\fontsize{\fsize}{#1\fsize}\selectfont}%
  \ifx\svgwidth\undefined%
    \setlength{\unitlength}{350bp}%
    \ifx\svgscale\undefined%
      \relax%
    \else%
      \setlength{\unitlength}{\unitlength * \real{\svgscale}}%
    \fi%
  \else%
    \setlength{\unitlength}{\svgwidth}%
  \fi%
  \global\let\svgwidth\undefined%
  \global\let\svgscale\undefined%
  \makeatother%
  \begin{picture}(1,0.4)%
    \lineheight{1}%
    \setlength\tabcolsep{0pt}%
    \put(0,0){\includegraphics[width=\unitlength,page=1]{architecture_overview.pdf}}%
    \put(0.12142857,0.24285715){\makebox(0,0)[t]{\lineheight{1.25}\smash{\begin{tabular}[t]{c}ADC\end{tabular}}}}%
    \put(0.04999999,0.32857143){\makebox(0,0)[lt]{\lineheight{1.25}\smash{\begin{tabular}[t]{l}\tiny{2x}\end{tabular}}}}%
    \put(0,0){\includegraphics[width=\unitlength,page=2]{architecture_overview.pdf}}%
    \put(0.29285711,0.26428572){\makebox(0,0)[t]{\lineheight{1.25}\smash{\begin{tabular}[t]{c}Input\\Mapping\end{tabular}}}}%
    \put(0,0){\includegraphics[width=\unitlength,page=3]{architecture_overview.pdf}}%
    \put(0.2142857,0.27142858){\makebox(0,0)[t]{\lineheight{1.25}\smash{\begin{tabular}[t]{c}\tiny{2}\end{tabular}}}}%
    \put(0,0){\includegraphics[width=\unitlength,page=4]{architecture_overview.pdf}}%
    \put(0.29285711,0.10714285){\makebox(0,0)[t]{\lineheight{1.25}\smash{\begin{tabular}[t]{c}Noise\\Gen.\end{tabular}}}}%
    \put(0,0){\includegraphics[width=\unitlength,page=5]{architecture_overview.pdf}}%
    \put(0.50714283,0.2){\makebox(0,0)[t]{\lineheight{1.25}\smash{\begin{tabular}[t]{c}State\\Processing\\Slices\end{tabular}}}}%
    \put(0,0){\includegraphics[width=\unitlength,page=6]{architecture_overview.pdf}}%
    \put(0.37142851,0.27142858){\makebox(0,0)[t]{\lineheight{1.25}\smash{\begin{tabular}[t]{c}\tiny{3}\end{tabular}}}}%
    \put(0.2825291,0.17519686){\makebox(0,0)[rt]{\lineheight{1.25}\smash{\begin{tabular}[t]{r}\tiny{3}\end{tabular}}}}%
    \put(0.40714282,0.33571428){\makebox(0,0)[lt]{\lineheight{1.25}\smash{\begin{tabular}[t]{l}\tiny{3x}\end{tabular}}}}%
    \put(0,0){\includegraphics[width=\unitlength,page=7]{architecture_overview.pdf}}%
    \put(0.71428571,0.26428572){\makebox(0,0)[t]{\lineheight{1.25}\smash{\begin{tabular}[t]{c}Output\\Mapping\end{tabular}}}}%
    \put(0,0){\includegraphics[width=\unitlength,page=8]{architecture_overview.pdf}}%
    \put(0.62142857,0.27142857){\makebox(0,0)[t]{\lineheight{1.25}\smash{\begin{tabular}[t]{c}\tiny{6}\end{tabular}}}}%
    \put(0.78571433,0.27142858){\makebox(0,0)[t]{\lineheight{1.25}\smash{\begin{tabular}[t]{c}\tiny{2}\end{tabular}}}}%
    \put(0,0){\includegraphics[width=\unitlength,page=9]{architecture_overview.pdf}}%
    \put(0.89285722,0.24285715){\makebox(0,0)[t]{\lineheight{1.25}\smash{\begin{tabular}[t]{c}DAC\end{tabular}}}}%
    \put(0,0){\includegraphics[width=\unitlength,page=10]{architecture_overview.pdf}}%
    \put(0.21428572,0.22857143){\makebox(0,0)[t]{\lineheight{1.25}\smash{\begin{tabular}[t]{c}$u_i$\end{tabular}}}}%
    \put(0,0){\includegraphics[width=\unitlength,page=11]{architecture_overview.pdf}}%
    \put(0.18571429,0.36428572){\makebox(0,0)[lt]{\lineheight{1.25}\smash{\begin{tabular}[t]{l}FPGA DSP System\end{tabular}}}}%
    \put(0.37857141,0.22857143){\makebox(0,0)[t]{\lineheight{1.25}\smash{\begin{tabular}[t]{c}$z_j$\end{tabular}}}}%
    \put(0.30616095,0.17003482){\makebox(0,0)[lt]{\lineheight{1.25}\smash{\begin{tabular}[t]{l}$\xi_j$\end{tabular}}}}%
    \put(0.61797368,0.21857143){\makebox(0,0)[t]{\lineheight{1.25}\smash{\begin{tabular}[t]{c}$q_j, \dot{q}_j$\end{tabular}}}}%
    \put(0.79285717,0.22857143){\makebox(0,0)[t]{\lineheight{1.25}\smash{\begin{tabular}[t]{c}$y_k$\end{tabular}}}}%
    \put(0.8214286,0.32857143){\makebox(0,0)[lt]{\lineheight{1.25}\smash{\begin{tabular}[t]{l}\tiny{2x}\end{tabular}}}}%
    \put(0,0){\includegraphics[width=\unitlength,page=12]{architecture_overview.pdf}}%
  \end{picture}%
\endgroup%

%% file: graphics/input_matrix_arch.pdf_tex
\begingroup%
  \makeatletter%
  \providecommand\color[2][]{%
    \errmessage{(Inkscape) Color is used for the text in Inkscape, but the package 'color.sty' is not loaded}%
    \renewcommand\color[2][]{}%
  }%
  \providecommand\transparent[1]{%
    \errmessage{(Inkscape) Transparency is used (non-zero) for the text in Inkscape, but the package 'transparent.sty' is not loaded}%
    \renewcommand\transparent[1]{}%
  }%
  \providecommand\rotatebox[2]{#2}%
  \newcommand*\fsize{\dimexpr\f@size pt\relax}%
  \newcommand*\lineheight[1]{\fontsize{\fsize}{#1\fsize}\selectfont}%
  \ifx\svgwidth\undefined%
    \setlength{\unitlength}{280bp}%
    \ifx\svgscale\undefined%
      \relax%
    \else%
      \setlength{\unitlength}{\unitlength * \real{\svgscale}}%
    \fi%
  \else%
    \setlength{\unitlength}{\svgwidth}%
  \fi%
  \global\let\svgwidth\undefined%
  \global\let\svgscale\undefined%
  \makeatother%
  \begin{picture}(1,0.75)%
    \lineheight{1}%
    \setlength\tabcolsep{0pt}%
    \put(0,0){\includegraphics[width=\unitlength,page=1]{input_matrix_arch.pdf}}%
    \put(0.24999999,0.38392858){\makebox(0,0)[t]{\lineheight{1.25}\smash{\begin{tabular}[t]{c}$\alpha_1(u_1[n])$\end{tabular}}}}%
    \put(0,0){\includegraphics[width=\unitlength,page=2]{input_matrix_arch.pdf}}%
    \put(0.06249996,0.38392858){\makebox(0,0)[rt]{\lineheight{1.25}\smash{\begin{tabular}[t]{r}$u_1[n]$\end{tabular}}}}%
    \put(0,0){\includegraphics[width=\unitlength,page=3]{input_matrix_arch.pdf}}%
    \put(0.24999997,0.27678574){\makebox(0,0)[t]{\lineheight{1.25}\smash{\begin{tabular}[t]{c}$\alpha_2(u_2[n])$\end{tabular}}}}%
    \put(0,0){\includegraphics[width=\unitlength,page=4]{input_matrix_arch.pdf}}%
    \put(0.06249996,0.27678574){\makebox(0,0)[rt]{\lineheight{1.25}\smash{\begin{tabular}[t]{r}$u_2[n]$\end{tabular}}}}%
    \put(0,0){\includegraphics[width=\unitlength,page=5]{input_matrix_arch.pdf}}%
    \put(0.57142855,0.6607143){\makebox(0,0)[t]{\lineheight{1.25}\smash{\begin{tabular}[t]{c}$a_{1,1}$\end{tabular}}}}%
    \put(0,0){\includegraphics[width=\unitlength,page=6]{input_matrix_arch.pdf}}%
    \put(0.57142855,0.55357144){\makebox(0,0)[t]{\lineheight{1.25}\smash{\begin{tabular}[t]{c}$a_{1,2}$\end{tabular}}}}%
    \put(0,0){\includegraphics[width=\unitlength,page=7]{input_matrix_arch.pdf}}%
    \put(0.69642861,0.55892859){\makebox(0,0)[t]{\lineheight{1.25}\smash{\begin{tabular}[t]{c}$\Sigma$\end{tabular}}}}%
    \put(0,0){\includegraphics[width=\unitlength,page=8]{input_matrix_arch.pdf}}%
    \put(0.94642852,0.56249999){\makebox(0,0)[lt]{\lineheight{1.25}\smash{\begin{tabular}[t]{l}$z_1[n]$\end{tabular}}}}%
    \put(0,0){\includegraphics[width=\unitlength,page=9]{input_matrix_arch.pdf}}%
    \put(0.57142855,0.42857145){\makebox(0,0)[t]{\lineheight{1.25}\smash{\begin{tabular}[t]{c}$a_{2,1}$\end{tabular}}}}%
    \put(0,0){\includegraphics[width=\unitlength,page=10]{input_matrix_arch.pdf}}%
    \put(0.57142855,0.32142863){\makebox(0,0)[t]{\lineheight{1.25}\smash{\begin{tabular}[t]{c}$a_{2,2}$\end{tabular}}}}%
    \put(0,0){\includegraphics[width=\unitlength,page=11]{input_matrix_arch.pdf}}%
    \put(0.69642861,0.32678577){\makebox(0,0)[t]{\lineheight{1.25}\smash{\begin{tabular}[t]{c}$\Sigma$\end{tabular}}}}%
    \put(0,0){\includegraphics[width=\unitlength,page=12]{input_matrix_arch.pdf}}%
    \put(0.57142855,0.1964286){\makebox(0,0)[t]{\lineheight{1.25}\smash{\begin{tabular}[t]{c}$a_{3,1}$\end{tabular}}}}%
    \put(0,0){\includegraphics[width=\unitlength,page=13]{input_matrix_arch.pdf}}%
    \put(0.57142855,0.08928568){\makebox(0,0)[t]{\lineheight{1.25}\smash{\begin{tabular}[t]{c}$a_{3,2}$\end{tabular}}}}%
    \put(0,0){\includegraphics[width=\unitlength,page=14]{input_matrix_arch.pdf}}%
    \put(0.69642861,0.09464282){\makebox(0,0)[t]{\lineheight{1.25}\smash{\begin{tabular}[t]{c}$\Sigma$\end{tabular}}}}%
    \put(0,0){\includegraphics[width=\unitlength,page=15]{input_matrix_arch.pdf}}%
    \put(0.83035716,0.55892859){\makebox(0,0)[t]{\lineheight{1.25}\smash{\begin{tabular}[t]{c}$\Sigma$\end{tabular}}}}%
    \put(0,0){\includegraphics[width=\unitlength,page=16]{input_matrix_arch.pdf}}%
    \put(0.57142855,0.6607143){\makebox(0,0)[t]{\lineheight{1.25}\smash{\begin{tabular}[t]{c}$a_{1,1}$\end{tabular}}}}%
    \put(0,0){\includegraphics[width=\unitlength,page=17]{input_matrix_arch.pdf}}%
    \put(0.85674158,0.64432066){\makebox(0,0)[lt]{\lineheight{1.25}\smash{\begin{tabular}[t]{l}$b_1$\end{tabular}}}}%
    \put(0.76785712,0.71428572){\makebox(0,0)[rt]{\lineheight{1.25}\smash{\begin{tabular}[t]{r}$\xi_1[n]$\end{tabular}}}}%
    \put(0,0){\includegraphics[width=\unitlength,page=18]{input_matrix_arch.pdf}}%
    \put(0.94642852,0.33035714){\makebox(0,0)[lt]{\lineheight{1.25}\smash{\begin{tabular}[t]{l}$z_2[n]$\end{tabular}}}}%
    \put(0,0){\includegraphics[width=\unitlength,page=19]{input_matrix_arch.pdf}}%
    \put(0.83035716,0.32678573){\makebox(0,0)[t]{\lineheight{1.25}\smash{\begin{tabular}[t]{c}$\Sigma$\end{tabular}}}}%
    \put(0,0){\includegraphics[width=\unitlength,page=20]{input_matrix_arch.pdf}}%
    \put(0.85674158,0.41217779){\makebox(0,0)[lt]{\lineheight{1.25}\smash{\begin{tabular}[t]{l}$b_2$\end{tabular}}}}%
    \put(0,0){\includegraphics[width=\unitlength,page=21]{input_matrix_arch.pdf}}%
    \put(0.94642845,0.09821423){\makebox(0,0)[lt]{\lineheight{1.25}\smash{\begin{tabular}[t]{l}$z_3[n]$\end{tabular}}}}%
    \put(0,0){\includegraphics[width=\unitlength,page=22]{input_matrix_arch.pdf}}%
    \put(0.83035708,0.09464282){\makebox(0,0)[t]{\lineheight{1.25}\smash{\begin{tabular}[t]{c}$\Sigma$\end{tabular}}}}%
    \put(0,0){\includegraphics[width=\unitlength,page=23]{input_matrix_arch.pdf}}%
    \put(0.8567415,0.18003495){\makebox(0,0)[lt]{\lineheight{1.25}\smash{\begin{tabular}[t]{l}$b_3$\end{tabular}}}}%
    \put(0,0){\includegraphics[width=\unitlength,page=24]{input_matrix_arch.pdf}}%
    \put(0.76785712,0.48214287){\makebox(0,0)[rt]{\lineheight{1.25}\smash{\begin{tabular}[t]{r}$\xi_2[n]$\end{tabular}}}}%
    \put(0,0){\includegraphics[width=\unitlength,page=25]{input_matrix_arch.pdf}}%
    \put(0.76785712,0.25){\makebox(0,0)[rt]{\lineheight{1.25}\smash{\begin{tabular}[t]{r}$\xi_3[n]$\end{tabular}}}}%
    \put(0,0){\includegraphics[width=\unitlength,page=26]{input_matrix_arch.pdf}}%
  \end{picture}%
\endgroup%

%% file: graphics/output_matrix_arch.pdf_tex
\begingroup%
  \makeatletter%
  \providecommand\color[2][]{%
    \errmessage{(Inkscape) Color is used for the text in Inkscape, but the package 'color.sty' is not loaded}%
    \renewcommand\color[2][]{}%
  }%
  \providecommand\transparent[1]{%
    \errmessage{(Inkscape) Transparency is used (non-zero) for the text in Inkscape, but the package 'transparent.sty' is not loaded}%
    \renewcommand\transparent[1]{}%
  }%
  \providecommand\rotatebox[2]{#2}%
  \newcommand*\fsize{\dimexpr\f@size pt\relax}%
  \newcommand*\lineheight[1]{\fontsize{\fsize}{#1\fsize}\selectfont}%
  \ifx\svgwidth\undefined%
    \setlength{\unitlength}{290bp}%
    \ifx\svgscale\undefined%
      \relax%
    \else%
      \setlength{\unitlength}{\unitlength * \real{\svgscale}}%
    \fi%
  \else%
    \setlength{\unitlength}{\svgwidth}%
  \fi%
  \global\let\svgwidth\undefined%
  \global\let\svgscale\undefined%
  \makeatother%
  \begin{picture}(1,0.84482759)%
    \lineheight{1}%
    \setlength\tabcolsep{0pt}%
    \put(0,0){\includegraphics[width=\unitlength,page=1]{output_matrix_arch.pdf}}%
    \put(0.53448272,0.77586205){\makebox(0,0)[t]{\lineheight{1.25}\smash{\begin{tabular}[t]{c}$\beta_{1,1}(x)$\end{tabular}}}}%
    \put(0,0){\includegraphics[width=\unitlength,page=2]{output_matrix_arch.pdf}}%
    \put(0.53448272,0.63793103){\makebox(0,0)[t]{\lineheight{1.25}\smash{\begin{tabular}[t]{c}$\beta_{2,1}(x)$\end{tabular}}}}%
    \put(0,0){\includegraphics[width=\unitlength,page=3]{output_matrix_arch.pdf}}%
    \put(0.53448272,0.5){\makebox(0,0)[t]{\lineheight{1.25}\smash{\begin{tabular}[t]{c}$\beta_{3,1}(x)$\end{tabular}}}}%
    \put(0,0){\includegraphics[width=\unitlength,page=4]{output_matrix_arch.pdf}}%
    \put(0.83620689,0.63448276){\makebox(0,0)[t]{\lineheight{1.25}\smash{\begin{tabular}[t]{c}$\Sigma$\end{tabular}}}}%
    \put(0,0){\includegraphics[width=\unitlength,page=5]{output_matrix_arch.pdf}}%
    \put(0.94827587,0.63793105){\makebox(0,0)[lt]{\lineheight{1.25}\smash{\begin{tabular}[t]{l}$y_1[n]$\end{tabular}}}}%
    \put(0,0){\includegraphics[width=\unitlength,page=6]{output_matrix_arch.pdf}}%
    \put(0.53448272,0.32758621){\makebox(0,0)[t]{\lineheight{1.25}\smash{\begin{tabular}[t]{c}$\beta_{1,2}(x)$\end{tabular}}}}%
    \put(0,0){\includegraphics[width=\unitlength,page=7]{output_matrix_arch.pdf}}%
    \put(0.53448272,0.18965519){\makebox(0,0)[t]{\lineheight{1.25}\smash{\begin{tabular}[t]{c}$\beta_{2,2}(x)$\end{tabular}}}}%
    \put(0,0){\includegraphics[width=\unitlength,page=8]{output_matrix_arch.pdf}}%
    \put(0.53448272,0.0517242){\makebox(0,0)[t]{\lineheight{1.25}\smash{\begin{tabular}[t]{c}$b_{3,2}(x)$\end{tabular}}}}%
    \put(0,0){\includegraphics[width=\unitlength,page=9]{output_matrix_arch.pdf}}%
    \put(0.83620689,0.18620693){\makebox(0,0)[t]{\lineheight{1.25}\smash{\begin{tabular}[t]{c}$\Sigma$\end{tabular}}}}%
    \put(0,0){\includegraphics[width=\unitlength,page=10]{output_matrix_arch.pdf}}%
    \put(0.94827587,0.18965519){\makebox(0,0)[lt]{\lineheight{1.25}\smash{\begin{tabular}[t]{l}$y_2[n]$\end{tabular}}}}%
    \put(0,0){\includegraphics[width=\unitlength,page=11]{output_matrix_arch.pdf}}%
    \put(0.06034483,0.80172414){\makebox(0,0)[rt]{\lineheight{1.25}\smash{\begin{tabular}[t]{r}$q_1[n]$\end{tabular}}}}%
    \put(0.06034483,0.74999999){\makebox(0,0)[rt]{\lineheight{1.25}\smash{\begin{tabular}[t]{r}$\dot{q}_1[n]$\end{tabular}}}}%
    \put(0.06034483,0.66379312){\makebox(0,0)[rt]{\lineheight{1.25}\smash{\begin{tabular}[t]{r}$q_2[n]$\end{tabular}}}}%
    \put(0.06034484,0.61206897){\makebox(0,0)[rt]{\lineheight{1.25}\smash{\begin{tabular}[t]{r}$\dot{q}_2[n]$\end{tabular}}}}%
    \put(0.06034483,0.5258621){\makebox(0,0)[rt]{\lineheight{1.25}\smash{\begin{tabular}[t]{r}$q_3[n]$\end{tabular}}}}%
    \put(0.06034484,0.47413795){\makebox(0,0)[rt]{\lineheight{1.25}\smash{\begin{tabular}[t]{r}$\dot{q}_3[n]$\end{tabular}}}}%
    \put(0,0){\includegraphics[width=\unitlength,page=12]{output_matrix_arch.pdf}}%
    \put(0.68965509,0.81896552){\makebox(0,0)[t]{\lineheight{1.25}\smash{\begin{tabular}[t]{c}$d_{1,1}$\end{tabular}}}}%
    \put(0,0){\includegraphics[width=\unitlength,page=13]{output_matrix_arch.pdf}}%
    \put(0.68965509,0.68103449){\makebox(0,0)[t]{\lineheight{1.25}\smash{\begin{tabular}[t]{c}$d_{1,2}$\end{tabular}}}}%
    \put(0,0){\includegraphics[width=\unitlength,page=14]{output_matrix_arch.pdf}}%
    \put(0.68965509,0.54310346){\makebox(0,0)[t]{\lineheight{1.25}\smash{\begin{tabular}[t]{c}$d_{1,3}$\end{tabular}}}}%
    \put(0,0){\includegraphics[width=\unitlength,page=15]{output_matrix_arch.pdf}}%
    \put(0.68965509,0.37068969){\makebox(0,0)[t]{\lineheight{1.25}\smash{\begin{tabular}[t]{c}$d_{2,1}$\end{tabular}}}}%
    \put(0,0){\includegraphics[width=\unitlength,page=16]{output_matrix_arch.pdf}}%
    \put(0.68965516,0.23275863){\makebox(0,0)[t]{\lineheight{1.25}\smash{\begin{tabular}[t]{c}$d_{2,2}$\end{tabular}}}}%
    \put(0,0){\includegraphics[width=\unitlength,page=17]{output_matrix_arch.pdf}}%
    \put(0.68965509,0.09482764){\makebox(0,0)[t]{\lineheight{1.25}\smash{\begin{tabular}[t]{c}$d_{2,3}$\end{tabular}}}}%
    \put(0,0){\includegraphics[width=\unitlength,page=18]{output_matrix_arch.pdf}}%
  \end{picture}%
\endgroup%

%% file: graphics/integrator_symplectic_lm2.pdf_tex
\begingroup%
  \makeatletter%
  \providecommand\color[2][]{%
    \errmessage{(Inkscape) Color is used for the text in Inkscape, but the package 'color.sty' is not loaded}%
    \renewcommand\color[2][]{}%
  }%
  \providecommand\transparent[1]{%
    \errmessage{(Inkscape) Transparency is used (non-zero) for the text in Inkscape, but the package 'transparent.sty' is not loaded}%
    \renewcommand\transparent[1]{}%
  }%
  \providecommand\rotatebox[2]{#2}%
  \newcommand*\fsize{\dimexpr\f@size pt\relax}%
  \newcommand*\lineheight[1]{\fontsize{\fsize}{#1\fsize}\selectfont}%
  \ifx\svgwidth\undefined%
    \setlength{\unitlength}{325bp}%
    \ifx\svgscale\undefined%
      \relax%
    \else%
      \setlength{\unitlength}{\unitlength * \real{\svgscale}}%
    \fi%
  \else%
    \setlength{\unitlength}{\svgwidth}%
  \fi%
  \global\let\svgwidth\undefined%
  \global\let\svgscale\undefined%
  \makeatother%
  \begin{picture}(1,0.46923077)%
    \lineheight{1}%
    \setlength\tabcolsep{0pt}%
    \put(0,0){\includegraphics[width=\unitlength,page=1]{integrator_symplectic_lm2.pdf}}%
    \put(0.33846158,0.40769229){\makebox(0,0)[t]{\lineheight{1.25}\smash{\begin{tabular}[t]{c}$\kappa$\end{tabular}}}}%
    \put(0,0){\includegraphics[width=\unitlength,page=2]{integrator_symplectic_lm2.pdf}}%
    \put(0.24615385,0.36615384){\makebox(0,0)[t]{\lineheight{1.25}\smash{\begin{tabular}[t]{c}$\Sigma$\end{tabular}}}}%
    \put(0,0){\includegraphics[width=\unitlength,page=3]{integrator_symplectic_lm2.pdf}}%
    \put(0.13845565,0.36773435){\makebox(0,0)[t]{\lineheight{1.25}\smash{\begin{tabular}[t]{c}$1/2$\end{tabular}}}}%
    \put(0,0){\includegraphics[width=\unitlength,page=4]{integrator_symplectic_lm2.pdf}}%
    \put(0.13849135,0.2906718){\makebox(0,0)[t]{\lineheight{1.25}\smash{\begin{tabular}[t]{c}$z^{-1}$\end{tabular}}}}%
    \put(0,0){\includegraphics[width=\unitlength,page=5]{integrator_symplectic_lm2.pdf}}%
    \put(0.25410829,0.34347428){\makebox(0,0)[lt]{\lineheight{1.25}\smash{\begin{tabular}[t]{l}\tiny{$-$}\end{tabular}}}}%
    \put(0,0){\includegraphics[width=\unitlength,page=6]{integrator_symplectic_lm2.pdf}}%
    \put(0.45384032,0.36773358){\makebox(0,0)[t]{\lineheight{1.25}\smash{\begin{tabular}[t]{c}$2^{\lambda}$\end{tabular}}}}%
    \put(0,0){\includegraphics[width=\unitlength,page=7]{integrator_symplectic_lm2.pdf}}%
    \put(0.54615388,0.36615384){\makebox(0,0)[t]{\lineheight{1.25}\smash{\begin{tabular}[t]{c}$\Sigma$\end{tabular}}}}%
    \put(0,0){\includegraphics[width=\unitlength,page=8]{integrator_symplectic_lm2.pdf}}%
    \put(0.60772214,0.29067178){\makebox(0,0)[t]{\lineheight{1.25}\smash{\begin{tabular}[t]{c}$z^{-1}$\end{tabular}}}}%
    \put(0,0){\includegraphics[width=\unitlength,page=9]{integrator_symplectic_lm2.pdf}}%
    \put(0.73845572,0.36773435){\makebox(0,0)[t]{\lineheight{1.25}\smash{\begin{tabular}[t]{c}$2^{\lambda_{\dot q}}$\end{tabular}}}}%
    \put(0,0){\includegraphics[width=\unitlength,page=10]{integrator_symplectic_lm2.pdf}}%
    \put(0.85384616,0.40769229){\makebox(0,0)[t]{\lineheight{1.25}\smash{\begin{tabular}[t]{c}$\kappa_{\dot q}$\end{tabular}}}}%
    \put(0,0){\includegraphics[width=\unitlength,page=11]{integrator_symplectic_lm2.pdf}}%
    \put(0.96560614,0.36745869){\makebox(0,0)[lt]{\lineheight{1.25}\smash{\begin{tabular}[t]{l}$\dot{q}[n+1]$\end{tabular}}}}%
    \put(0.03275055,0.36692052){\makebox(0,0)[rt]{\lineheight{1.25}\smash{\begin{tabular}[t]{r}$\ddot{q}[n]$\end{tabular}}}}%
    \put(0,0){\includegraphics[width=\unitlength,page=12]{integrator_symplectic_lm2.pdf}}%
    \put(0.13849135,0.05990255){\makebox(0,0)[t]{\lineheight{1.25}\smash{\begin{tabular}[t]{c}$z^{-1}$\end{tabular}}}}%
    \put(0,0){\includegraphics[width=\unitlength,page=13]{integrator_symplectic_lm2.pdf}}%
    \put(0.33846158,0.19230768){\makebox(0,0)[t]{\lineheight{1.25}\smash{\begin{tabular}[t]{c}$\kappa$\end{tabular}}}}%
    \put(0,0){\includegraphics[width=\unitlength,page=14]{integrator_symplectic_lm2.pdf}}%
    \put(0.45384032,0.15234897){\makebox(0,0)[t]{\lineheight{1.25}\smash{\begin{tabular}[t]{c}$2^{\lambda}$\end{tabular}}}}%
    \put(0,0){\includegraphics[width=\unitlength,page=15]{integrator_symplectic_lm2.pdf}}%
    \put(0.13846153,0.15076923){\makebox(0,0)[t]{\lineheight{1.25}\smash{\begin{tabular}[t]{c}$\Sigma$\end{tabular}}}}%
    \put(0,0){\includegraphics[width=\unitlength,page=16]{integrator_symplectic_lm2.pdf}}%
    \put(0.54615382,0.15076923){\makebox(0,0)[t]{\lineheight{1.25}\smash{\begin{tabular}[t]{c}$\Sigma$\end{tabular}}}}%
    \put(0,0){\includegraphics[width=\unitlength,page=17]{integrator_symplectic_lm2.pdf}}%
    \put(0.60772221,0.07528713){\makebox(0,0)[t]{\lineheight{1.25}\smash{\begin{tabular}[t]{c}$z^{-1}$\end{tabular}}}}%
    \put(0,0){\includegraphics[width=\unitlength,page=18]{integrator_symplectic_lm2.pdf}}%
    \put(0.73845565,0.15234973){\makebox(0,0)[t]{\lineheight{1.25}\smash{\begin{tabular}[t]{c}$2^{\lambda_q}$\end{tabular}}}}%
    \put(0,0){\includegraphics[width=\unitlength,page=19]{integrator_symplectic_lm2.pdf}}%
    \put(0.85384616,0.19230768){\makebox(0,0)[t]{\lineheight{1.25}\smash{\begin{tabular}[t]{c}$\kappa_q$\end{tabular}}}}%
    \put(0,0){\includegraphics[width=\unitlength,page=20]{integrator_symplectic_lm2.pdf}}%
    \put(0.96560614,0.15207404){\makebox(0,0)[lt]{\lineheight{1.25}\smash{\begin{tabular}[t]{l}$q[n+1]$\end{tabular}}}}%
    \put(0,0){\includegraphics[width=\unitlength,page=21]{integrator_symplectic_lm2.pdf}}%
    \put(0.23076336,0.1523497){\makebox(0,0)[t]{\lineheight{1.25}\smash{\begin{tabular}[t]{c}$1/2$\end{tabular}}}}%
    \put(0,0){\includegraphics[width=\unitlength,page=22]{integrator_symplectic_lm2.pdf}}%
    \put(0.39999586,0.43804259){\makebox(0,0)[t]{\lineheight{1.25}\smash{\begin{tabular}[t]{c}$t_s$ scaling\end{tabular}}}}%
    \put(0.59625657,0.43804259){\makebox(0,0)[t]{\lineheight{1.25}\smash{\begin{tabular}[t]{c}Accumulator\end{tabular}}}}%
    \put(0.79261605,0.43804259){\makebox(0,0)[t]{\lineheight{1.25}\smash{\begin{tabular}[t]{c}Rescaling\end{tabular}}}}%
    \put(0,0){\includegraphics[width=\unitlength,page=23]{integrator_symplectic_lm2.pdf}}%
  \end{picture}%
\endgroup%

%% file: pages/simulator/model_config_setup.tex
\subsection{Model Configuration}
\label{sec:model_configuration}
The HIL simulator enables users to configure custom models through a MATLAB-based API, which translates user-defined continuous-time state-space equations into hardware-compatible configurations. The MATLAB API automatically handles renormalization, scaling, and optimization based on user-provided simulation scenarios, ensuring numerical stability and hardware compatibility without requiring manual tuning.

A lightweight configuration server, implemented in C, runs on the processor subsystem of the SoC and receives configuration data from the MATLAB application. It converts this data into the required binary format and transfers it to the FPGA IP core via memory-mapped I/O. In this way, MATLAB provides a user-friendly front end for model setup, while the server handles direct interaction with the hardware.

\subsection{Setup}
\label{subsec:sim_setup}
The prerequisites for running the Simulator are:
\begin{itemize}
\item Red Pitaya STEMlab 125-14 Z7020 (Gen 1 or 2) (referred to as RP)
\item MATLAB 2021b or newer (required for model configuration and simulation).
\item Vitis Model Composer 2021b (optional, only needed for hardware-accurate simulations using the IP core model).
\item \textbf{Optional:} Internet access for the RP.
\item \textbf{Optional:} Oscilloscope and signal generator for I/O calibration.
\end{itemize}

\subsubsection{Hardware}
\label{subsubsec:sim_hardware_setup}

\begin{enumerate}
    \item Ensure the RP is accessible via SSH and SCP. We recommend using WinSCP and PuTTY for Windows.

    \item Transfer the distribution package \texttt{package/redpitaya\_hil\_sim\_package.tar.gz} to the RP using SCP.

    \item Extract the package on the RP using: \\
        \texttt{tar -xf redpitaya\_hil\_sim\_package.tar.gz}

    \item Run \texttt{./start.sh} to initialize the FPGA bitstream and start the TCP server, which then starts listening on port \texttt{1000}.
    
\end{enumerate}

\subsubsection{Host PC setup}
\label{subsubsec:sim_sw_setup}
\begin{enumerate}
    \item Clone the HIL simulator repository to your host PC or copy only the MATLAB library from \texttt{/matlab/hilsim\_lib} to your MATLAB working directory and add it to the MATLAB path using \texttt{addpath('hilsim\_lib')}.
    \item Execute the provided demo configuration script \texttt{hilsim\_2d\_saddle\_example.m}, or follow the next section.
\end{enumerate}

\subsubsection{Calibration}
The MATLAB interface allows the user to calibrate the ADC and DAC channels to account for offset and gain variations using the provided script \texttt{matlab\_calibration.m}. This script performs a two-point calibration by prompting the user to connect known DC voltages and measure outputs, saving the results to \texttt{calib.mat} for later use with the HIL simulator client.

%% file: pages/example.tex
\section{Example: Particle in a 2D saddle potential}
\label{sec:example}

This section demonstrates the configuration and deployment of a levitated-particle model for HIL testing of a feedback-cooling controller. The example is inspired by our experimental work presented in \cite{mlynarFeedbackStabilizationNanoparticle2026} and was selected because the saddle-potential configuration contains an unstable direction and therefore provides a demanding benchmark for both the simulator and the LQG controller. The model simulates a particle exploring an adjustable optical double-well potential in two degrees of freedom. Due to measurement cross-talk, both detection channels contain information about the displacements of both axes. Forces can be exerted on the particle through analog inputs, which exhibit weak coupling to the other axes. Additionally, the particle is excited by a stochastic force that simulates the thermal bath. The optical potential $V$ can be switched between the confining potential $V_\mathrm{con}$ and the saddle-potential configuration $V_\mathrm{sad}$.

The plant dynamics are given by the stochastic differential equations of motion:
\begin{subequations}
\begin{gather}
\ddot{q}_1(t) = -\frac{\gamma}{m} \dot{q}_1(t) - \frac{1}{m}\frac{\partial V(q_1,q_2)}{\partial q_1} + \frac{1}{m} \left( C_{f1,1} u_1(t) + C_{f1,2} u_2(t) + \sqrt{P_w} \xi(t) \right), \\
\ddot{q}_2(t) = -\frac{\gamma}{m} \dot{q}_2(t) - \frac{1}{m}\frac{\partial V(q_1,q_2)}{\partial q_2} + \frac{1}{m} \left( C_{f2,1} u_1(t) + C_{f2,2} u_2(t) + \sqrt{P_w} \xi(t) \right),\\
V_\mathrm{con}(q_1, q_2) = 0.5\Omega_1^2 q_1^2 + 0.5\Omega_2^2 q_2^2,\\
V_\mathrm{sad}(q_1, q_2) = 0.25\alpha_1^4 q_1^4 - 0.5\Omega_1^2 q_1^2 + 0.5\Omega_2^2 q_2^2.
\end{gather}

The potential $V(q_1, q_2)$ is separable in $q_1$ and $q_2$, which makes it directly compatible with the simulator architecture described in Section \ref{subsec:sim_model}. The output function is defined as
\begin{align} 
    y_1(t) =& C_{e1,1} q_1(t) + C_{e1,2} q_2(t) \\
    y_2(t) =& C_{e2,1} q_1(t) + C_{e2,2} q_2(t).
\end{align}
\end{subequations}

The parameters are summarized in Table \ref{tab:example_parameters}.
\begin{table}[ht]
\centering
\def\arraystretch{1.3}
\begin{tabular}{|l|c|c|}
\hline
\textbf{Parameter} & \textbf{Symbol} & \textbf{Value} \\
\hline
\hline
Mass of the particle & $m$ & \SI{5e-17}{\kilogram} \\
\hline
Quadratic component of the $q_1$ potential & $\Omega_1$ & $2\pi\cross \SI{15}{\kilo\hertz}$ \\
\hline
Quartic component of the $q_1$ potential & $\alpha_1$ & $2\pi\cross \SI{40}{\kilo\hertz}$ \\
\hline
Natural frequency of the $q_2$ axis & $\Omega_2$ & $2\pi\cross \SI{35}{\kilo\hertz}$ \\
\hline
Damping rate & $\Gamma$ & $2\pi\cross \SI{1}{\kilo\hertz}$ \\
\hline
Power of the process noise & $P_w$ & $\SI{3.2e-34}{\newton^2\per\second^2}$ \\
\hline
Calibration coefficient of $q_1$ in $y_1$ & $C_{e1,1}$ & $\SI{4e6}{\volt\per\meter}$\\
\hline
Calibration coefficient of $q_1$ in $y_2$ & $C_{e1,2}$ & $\SI{1e5}{\volt\per\meter}$ \\
\hline
Calibration coefficient of $q_2$ in $y_1$ & $C_{e2,1}$ & $\SI{2e5}{\volt\per\meter}$ \\
\hline
Calibration coefficient of $q_2$ in $y_2$ & $C_{e2,2}$ & $\SI{8e6}{\volt\per\meter}$ \\
\hline
Force coupling coefficient of $u_1$ in $\dot{q}_1$ & $C_{f1,1}$ & $\SI{5e-12}{\newton\per\volt}$ \\
\hline
Force coupling coefficient of $u_2$ in $\dot{q}_1$ & $C_{f1,2}$ & $\SI{1e-12}{\newton\per\volt}$ \\
\hline
Force coupling coefficient of $u_1$ in $\dot{q}_2$ & $C_{f2,1}$ & $\SI{1e-12}{\newton\per\volt}$ \\
\hline
Force coupling coefficient of $u_2$ in $\dot{q}_2$ & $C_{f2,2}$ & $\SI{5e-12}{\newton\per\volt}$ \\
\hline
\end{tabular}
\caption{Parameters of the plant dynamics and output function.}
\label{tab:example_parameters}
\end{table}

First, we configure the HIL Simulator and then, in the next section, the LQG controller. It should be noted that fully realizing this example requires two Red Pitaya boards.

\subsection{Simulator: Plant configuration}\label{subsec:example_sim}
This section follows the script \texttt{hilsim\_2d\_saddle\_example.m}. First, define the model parameters and the potential functions in MATLAB:

\begin{minted}{matlab}
%% Particle Model Parameters & Potential Function
p.m = 5e-17;             % Particle mass [kg]
p.C_f1_1 = 5e-12;        % Force coupling coefficient of u1 in q1 [N/V]
p.C_f1_2 = 1e-12;        % Force coupling coefficient of u2 in q1 [N/V]
p.C_f2_1 = 1e-12;        % Force coupling coefficient of u1 in q2 [N/V]
p.C_f2_2 = 5e-12;        % Force coupling coefficient of u2 in q2 [N/V]
p.C_e1_1 = 4e6;          % Calibration coefficient of q1 in y1 [V/m]
p.C_e1_2 = 1e5;          % Calibration coefficient of q2 in y1 [V/m]
p.C_e2_1 = 2e5;          % Calibration coefficient of q1 in y2 [V/m]
p.C_e2_2 = 4e6;          % Calibration coefficient of q2 in y2 [V/m]
p.Omega_1 = 2*pi*15e3;   % Quadratic component of q1 potential [rad/s]
p.alpha_1 = 2*pi*40e3;   % Quartic component of q1 potential [rad/s]
p.Omega_2 = 2*pi*35e3;   % Natural frequency of q2 axis [rad/s]
p.Gamma = 2*pi*1e3;      % Damping rate [rad/s]
p.P_w = 3.2e-34;         % Process noise power [N^2/s^2]
% Potential = 0.25*alpha^4*q^4 - 0.5*Omega^2*q^2
p.force_func_q1     = @(q) q;
p.force_func_q1_alt = @(q) (p.alpha_1^4)/(p.Omega_1^2)*q.^3 - q;
\end{minted}

Note that the force function in the $q_1$ axis requires a scaling factor of $\Omega_1^2$ configured for the state slice, as shown in the next code block. We now configure the \texttt{HILSimModel} object:

\begin{minted}{matlab}
model = HILSimModel();
model.noise_sources(1).set_noise_power(p.P_w);
model.noise_sources(2).set_noise_power(p.P_w);
model.noise_sources(3).set_noise_power(p.P_w);
model.input_matrix.set_matrix([p.C_f1_1, p.C_f1_2, 1; ...
                               p.C_f2_1, p.C_f2_2, 1; ...
                               0, 0, 1]);
model.output_slices(1).linear_factors = [p.C_e1_1, p.C_e1_2, 0];
model.output_slices(2).linear_factors = [p.C_e2_1, p.C_e2_2, 0];

model.slices(1).u_nlf.set_bypass();
model.slices(1).u_factor = 1 / p.m;
model.slices(1).x_nlf.set_func(p.force_func_q1);
model.slices(1).x_nlf.set_func_alternate(p.force_func_q1_alt);
model.slices(1).x_factor = -p.Omega_1^2;
model.slices(1).xdot_nlf.set_bypass();
model.slices(1).xdot_factor = -p.Gamma;

model.slices(2).u_nlf.set_bypass();
model.slices(2).u_factor = 1 / p.m;
model.slices(2).x_factor = -p.Omega_2^2;
model.slices(2).x_nlf.set_bypass();
model.slices(2).xdot_nlf.set_bypass();
model.slices(2).xdot_factor = -p.Gamma;

model.slices(3).u_nlf.set_bypass();
model.slices(3).u_factor = 1 / p.m;
model.slices(3).x_nlf.set_bypass();
model.slices(3).x_factor = -1;
model.slices(3).xdot_nlf.set_bypass();
model.slices(3).xdot_factor = -p.Gamma;
\end{minted}

Then, the simulation scenarios must be defined for the algorithm to determine the appropriate numerical scaling. The scenarios should mimic the real inputs as closely as possible and should include scenarios for both potential variants if the force function switching is to be used.

\begin{minted}{matlab}
t_s = 1e-7;
num_samples = 1000*3;
sim_time = t_s * num_samples;

in_signals = zeros(2,2,num_samples);
in_signals(1,2,:) = linspace(0, t_s*(num_samples-1), num_samples);
in_signals(2,2,:) = linspace(0, t_s*(num_samples-1), num_samples);
for i=1:num_samples
    if i > 100 && i < 200
        in_signals(1,1,i) = 0.05;
        in_signals(2,1,i) = -0.05;
    end
end
scenario_0 = HILSimScenario(in_signals, sim_time);

in_signals = zeros(2,2,num_samples);
in_signals(1,2,:) = linspace(0, t_s*(num_samples-1), num_samples);
in_signals(2,2,:) = linspace(0, t_s*(num_samples-1), num_samples);
for i=1:num_samples
    if i > 100 && i < 352
        in_signals(1,1,i) = 0.02;
        in_signals(2,1,i) = -0.02;
    elseif i >= 352 && i <= 778
        in_signals(1,1,i) = -0.02;
        in_signals(2,1,i) = 0.02;
    end
end

scenario_1 = HILSimScenario(in_signals, sim_time);
scenario_1_alt = copy(scenario_1);
scenario_1_alt.nlf_alternate_switch.slice0.x_nlf = 1;
scenario_1_alt.nlf_alternate_switch.slice1.x_nlf = 1;
scenario_1_alt.nlf_alternate_switch.slice2.x_nlf = 1;

scenarios = [scenario_0, scenario_1, scenario_1_alt];
\end{minted}

Next, compile the model and simulate different implementation variants for verification. Verification based on visual overlap of the system outputs over time is rather coarse and depends on the configured scenarios. Significant mismatch can occur with the noise source enabled when comparing the \texttt{TD-FP} (time-discrete, fixed-point) model with the other variants. For \texttt{TD-FP}, the noise realization differs from \texttt{TC-DBL} (time-continuous, double) and \texttt{TD-DBL} (time-discrete, double), which can cause significant differences.

\begin{minted}{matlab}
model.compile(scenarios, false);
enable_noise = true;

sig_out_tc = model.sim_tc(scenarios(1), enable_noise);
sig_out_td = model.sim_td(scenarios(1), enable_noise);
sig_out_tdfp = model.sim_tdfp(scenarios(1), enable_noise);

figure();
axis tight;
hold on;
plot(sig_out_tc(1).Time, sig_out_tc(1).Data, "DisplayName", "TC-DBL");
plot(sig_out_td(1).Time, sig_out_td(1).Data, "DisplayName", "TD-DBL");
plot(sig_out_tdfp(1).Time, sig_out_tdfp(1).Data, "DisplayName", "TD-FP");
legend();
\end{minted}

Finally, load calibration and deploy to the Red Pitaya:

\begin{minted}{matlab}
client = HILSimClient(RP_IPADDR, RP_PORT);
client.connect();

% Send the compiled model to the RedPitaya
client.set_model(model);
client.send_config_update();

% Load calibration values from the saved calib data.
client.set_adc_calib(NaN, calib.adc0_factor, NaN, calib.adc1_factor);
client.set_dac_calib(calib.dac0_offset, calib.dac0_factor, calib.dac1_offset, calib.dac1_factor);

% Calibrate away any zero DC-offset in the ADC-inputs. This must be done *after*
% setting the rest of the ADC calibration values.
client.do_input_zero_calib();

% Setup external NLF switch for slice0/x
client.set_nlfsw_iomask(0x02);
client.set_nlfsw_sel(0x00);

% Reset all integrators in the model & start the simulation.
client.reset_simulation();
client.start_simulation();
\end{minted}

Now the simulation should be running, and you should be able to observe signals at the output. To switch the simulated potential branch from the confining profile to the saddle profile (and back), use:

\begin{minted}{matlab}
% Enable software NLF switching for x_nlf on slice 0
client.set_nlfsw_iomask(hex2dec('00'));

% Select primary NLF table (confining potential)
client.set_nlfsw_sel(hex2dec('00'));

% Select alternate NLF table (saddle potential)
client.set_nlfsw_sel(hex2dec('02'));
\end{minted}

To stop the simulation, execute the following code.

\begin{minted}{matlab}
% Stop the simulation.
client.stop_simulation();
client.reset_simulation();

% Disconnect from the RedPitaya.
client.disconnect();
\end{minted}

\subsection{LQG: Controller operation}
If you wish to test the LQG with the HIL Simulator, set up the Simulator as described in Section \ref{subsec:example_sim} and configure it first with the fully confining potential $V_\mathrm{con}$. Note that two controller parameter sets are used throughout the example. Set 1 is configured for the confining potential $V_\mathrm{con}$, while Set 2 is designed for operation in the saddle-potential configuration $V_\mathrm{sad}$.
 
Make sure your Red Pitaya is set up properly, according to Section \ref{subsec:lqg_setup}. Then connect the fast analog outputs of the LQG to the fast analog inputs of the Simulator, and vice versa, and make sure the server is running on the LQG. Open the MATLAB GUI, enter the IP address of your Red Pitaya, and click the \textbf{Click to connect} button. If the connection is successful, the button will turn green. You can then proceed with the configuration.

\begin{mdframed}
\textbf{Note:} If you do not have a second board available to use both the LQG and the Simulator simultaneously, you can either use a signal generator supplying sine waves at a given frequency, which removes the ability to verify the feedback action, or use the LQG Demo mode. To enable Demo mode, open the GUI and toggle the \textbf{Demo mode} switch (no hardware connection is required), then follow the rest of this section.
\end{mdframed}

Switch to the \textbf{Model configuration} tab, click the \textbf{Load parameters} button, and select the file \texttt{lqg\_2d\_saddle\_parameter\_set\_1.m} in the file explorer. When prompted, click \textbf{Set~1}, to push the parameters into the \textbf{Parameter set~1} tab. Repeat this process for the second parameter set. Feel free to inspect the parameters; they should correspond to Table \ref{tab:example_parameters}. Make sure the \textbf{Parameter set select} switch is set to \textbf{Set~1} and the \textbf{Enable feedback} switch is disabled, then click the \textbf{Push parameters} button to configure the FPGA with the estimator.

Switch to the \textbf{Data Acquisition} tab and uncheck Z, dZ, and Phi in the \textbf{Record} column of the signal selection table, as those signals are not used for the 2D configuration. Now press the \textbf{Record data} button, and after a short while, you should see data displayed in the plot window. To inspect the data in more detail, hover the mouse pointer over the plot axes, which reveals the plot inspection toolbar at the top of the axes. Check whether the X, dX, Y, and dY signals are approximately zero mean. If not, verify that the ADC1 and ADC2 signals are zero mean, and adjust the ADC offset accordingly in \textbf{Model configuration $\rightarrow$ IO calibration $\rightarrow$ Input calibration $\rightarrow$ Offset}.

To enable the feedback signal, set the \textbf{Enable feedback} switch to the \textbf{On} position. Flipping the switch will automatically push the parameters, so you do not need to update them manually. Record fresh data, and you should notice the overall decrease in variance in ADC1 and ADC2 (or lowering of the spectral peaks in the PSD mode). 

\begin{figure}[H]
    \centering
    \includegraphics[width=1\linewidth]{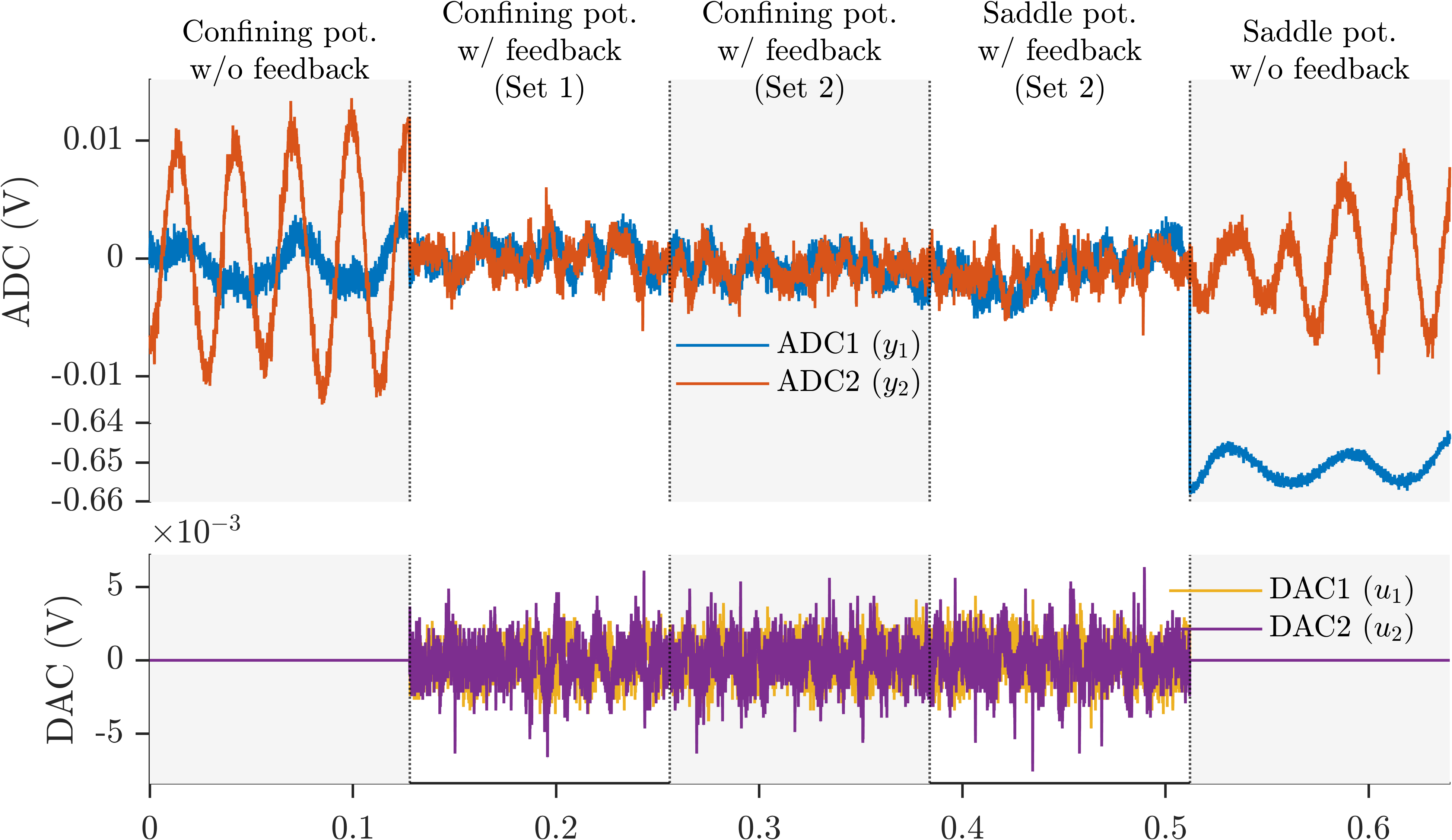}
    \caption{\textbf{Potential and controller switching protocol demonstration.} The upper axes show the measured outputs of the HIL Simulator, and the lower axis shows the feedback applied by the LQG. The shaded sections highlight the phases of the protocol. For clarity, the individual phases do not show one continuous-time recording; they show steady-state behavior for each configuration due to data-recording limitations. The protocol begins with a particle exploring the confining potential without feedback. The feedback is then enabled with \textbf{Set~1} (with $\Omega_1 = 2\pi\cross \SI{15}{\kilo\hertz}$), stabilizing both DOFs of the system. Next, to transition to the saddle potential, the LQG is switched to \textbf{Set~2} (with $\Omega_1 = -2\pi\cross \SI{15}{\kilo\hertz}$), followed by switching the HIL Simulator potential to the saddle potential. This figure illustrates that the particle remains confined throughout the transition. Finally, in the last phase, feedback is disabled, and the particle falls down the potential well. The example demonstrates stable closed-loop operation in both potential configurations and during the transition between them.}
    \label{fig:example_data_traces}
\end{figure}

To test controller performance in the saddle potential, first switch \textbf{Parameter set select} to \textbf{Set~2} while keeping feedback enabled, and then switch the potential in the Simulator using:

\begin{minted}{matlab}
% Switch simulator potential to saddle branch
client.set_nlfsw_sel(hex2dec('02'));
\end{minted}

During this transition, the controller is intentionally switched before the simulator potential. This ensures that the feedback law is already configured for the unstable saddle-potential dynamics when the potential is activated. Consequently, the particle remains confined throughout the transition, as illustrated in Figure \ref{fig:example_data_traces}.

We invite the user to try the opposite sequence by switching the Simulator first and only then the LQG. You should observe that the LQG, configured with the \textbf{Set~1} parameters and acting on the saddle potential $V_\mathrm{sad}$, is unable to compensate for the additional repulsive force of the potential.

If anything goes wrong during operation and you see unexpected behavior in the states, it is possible that the internal states have overflowed. To reset the states of the Simulator, first disengage the feedback signal originating from the LQG and then execute the following commands:
\begin{minted}{matlab}
client.reset_simulation();
client.start_simulation();
\end{minted}
To reset the states of the LQG, navigate to the \textbf{Connect} tab and press the \textbf{Reset internal states} button.